\documentclass{aa}  

\usepackage{graphicx}
\usepackage{txfonts}
\usepackage{natbib}

\usepackage{subcaption}
\usepackage{multicol}
\usepackage{afterpage}

\usepackage{xcolor}

\usepackage{float}

\usepackage{placeins}

\usepackage{tablefootnote}

\begin{document}

   \title{The mass distribution of stellar mergers: }

   \subtitle{A new scenario for several FS~CMa stars}

   \author{N.~Dvo\v{r}\'{a}kov\'{a} \inst{1}, D. Kor\v{c}\'akov\'a\inst{1}, F. Dinnbier \inst{1} \inst{\ref{inst_Jap}}
          \and
          P. Kroupa \inst{1} \inst{\ref{inst_Bonn}} 
          }

   \institute{Charles University, Faculty of Mathematics and Physics, Astronomical Institute, V Hole\v{s}ovi\v{c}k\'ach 2, CZ-180 00 Praha 8, Czech Republic\\
              \email{dvorakova.nela@seznam.cz}
    \and Helmholtz-Institut f\"{u}r Strahlen- und Kernphysik, University of Bonn, Nussallee 14-16, D-53115 Bonn, Germany\label{inst_Bonn}
         \and
             Department of Astronomy, Graduate School of Science, The University of Tokyo, 7-3-1 Hongo, Bunkyo-ku, Tokyo 113-0033, Japan \label{inst_Jap} \\
             }

   \date{Received ; accepted }

 
  \abstract
   {FS~CMa stars belong to a~diverse group of stars exhibiting the B[e]~phenomenon, which manifests itself mainly by the presence of forbidden emission lines and a~strong infrared (IR) excess in their spectra. Only a~few tens of FS~CMa stars are known and their nature is still unclear. Recently, a~strong magnetic field has been discovered in the FS CMa star IRAS~17449+2320. Its strength combined with an unusually high space velocity in the direction of the Galactic north pole point to the object having a post-merger nature. Such stellar mergers may provide an explanation for the complex and sometimes chaotic behaviour of some of the FS~CMa stars.}
   {In order to find out whether B-type stellar mergers are detectable, we did a~statistical study of numerical simulations using Aarseth's \textsc{nbody6} code. We show the importance of stellar mergers of low- to intermediate-mass stars (from $\approx$ 1.4 to $\approx$ 8~$\text{M}_{\odot}$) and for B-type stars in particular.}
   {We analysed two sets of N-body simulations with different initial orbital period distributions.
   In the simulations, more massive binaries are treated differently than less massive binaries and the mass limit usually used is 5~$\text{M}_{\odot}$. In addition to this, we also used the value of 2~$\text{M}_{\odot}$ to test the influence of this ambiguous limit on the results. Looking at mass, distance from their birth cluster, and velocity distributions, we investigated the statistical significance of individual spectral types in terms of merger dynamics and how merger events affect the stellar evolution.}
   {We have found that around 50~\% of stars in the simulated open clusters involved in the formation of mergers are B-type stars. As a~result, more than 50~\% of the merger products end up as a~B-type star as well. Also, between 12.54~\% and 23.24~\% of all B-type stars are mergers. These results are a~natural consequence of the initial mass function, initial distribution of the binary star parameters, and large range of masses for B-type stars. A~non-negligible fraction of mergers occurred before entering the common envelope phase and we detected merger events at extragalactic distances. The total amount of detected mergers could have an impact on the chemical evolution of galaxies. The resulting mass distribution of merger products shows a~peak for A-type stars, which is in agreement with observed massive Ap~stars. Post-mergers among late B-type FS~CMa stars could be the progenitors of very massive magnetic Ap~stars. Our results could also help to explain the nature of some magnetic white dwarfs. We present a~comparison of the W component and the space velocity of the simulated mergers with a~sample of observed FS~CMa stars.
   
   }
   {}


   \maketitle
%

\section{Introduction}
FS~CMa stars are a~subgroup of B[e]~type stars. These were first found by \cite{AllenSwings1976}, who noticed a~correlation between a~strong infrared (IR) excess and the presence of emission spectral lines in objects from a~large survey. \cite{Lamers1998} described this heterogeneous group and assigned observational criteria for the five classes of different objects exhibiting the B[e]~phenomenon: B[e]~supergiants (sgB[e]), pre-main-sequence stars (HerbigAe/B[e]), compact planetary nebulae B[e]-type stars (cPNB[e]), symbiotic B[e]~stars (SymB[e]), and the rest of the B[e]-type objects. These objects were either impossible to fit into any of these classes or would satisfy conditions for more than one.

\cite{Miroshnichenko_2007} studied the unclassified B[e]-type further and found that there are similarities for at least half of the known sample. These stars became a part of a~newly defined group of so-called FS~CMa stars. He described the observational criteria for this group. A~star needs to show strong emission of Balmer lines as well as emission lines of Fe~II and forbidden emission lines such as [Fe~II] or sometimes even [O~III]. A~large IR excess is another important feature and in the case of FS~CMa stars one should observe a peak at 10 - 30~$\mu$m with a~sharp decrease towards longer wavelengths. These stars are not within star-forming regions and if there is a~companion it should be cooler and fainter than the primary.

From an observational point of view, it is clear that we are dealing with complex objects heavily obscured by circumstellar dust and gas. Double-peaked and often asymmetrical profiles of emission lines point to the presence of a~gaseous disc around the central star. In some cases, there is a~small emission hump moving across the profile of a~Balmer line, indicating for example a~rotating structure around the star, as is described by \cite{Jerabkova2016} for the star HD~50138. In the case of MWC~342, radial velocities of the observed hump in the~H$\alpha$ line were explained with an~outflowing structure by \cite{Kucerova2013}. In several FS~CMa stars, discrete absorption features have been observed in the Na D1, D2 resonance doublet caused by material infall or ejecta \citep{Korcakova2022}. Among the most prominent forbidden lines are lines of [O~I] and sometimes we observe [N~II] and [S~II] lines. The unique atomic structure in some of the forbidden lines is very helpful in estimating important physical properties as they allow us to use nebular diagnostics. In the ultraviolet part of their spectra, the FS~CMa stars exhibit a~large number of absorption lines of iron group elements, the so-called ‘iron curtain’. These lines affect the visible and IR part of the spectrum where the absorbed energy is being reemitted. This spectral feature is also observed for classical novae as well as for symbiotic stars \citep{Shore1993}. FS~CMa stars are variable on many different timescales. Absorption lines can change from night to night, permitted emission lines vary on timescales from weeks to months, and the forbidden emission lines can vary on timescales of up to years \citep{Timescales2017}. Even photometric observations show variability. Due to a~non-homogeneous distribution of material in the circumstellar envelope (dusty discs or clumps of matter etc.), multi-periodic behaviour has been detected for some FS~CMa stars (e.g. \citealt{Goranskij2009} for CI~Cam, \citealt{Maravelias2018} for 3~Pup).

Our current understanding of the nature of FS~CMa stars is still incomplete. Among the suggestions to explain their observed properties have been Herbig Ae/Be type objects, but FS~CMa stars are isolated from star-forming regions, where these objects should be found. Several FS~CMa stars have been added to a~catalogue of post- asymptotic giant branch (post-AGB) stars or candidates \citep{Szczerba2007} but their IR excess peaks are at a~different wavelength and they are more dynamical than most post-AGB stars. 
\cite{Miroshnichenko_2007} discussed two cases of FS CMa stars with a~high mass-loss rate.
This mass-loss rate is an~order of a~magnitude higher than that of a~single star. An ongoing or recently terminated mass transfer between components of a~binary was suggested as an~explanation. Current models, however, work with the~assumption of free expansion of material, which is not observed \citep{Kucerova2013}.

The~discovery of a~strong magnetic field (6.2 $\pm$ 0.2 kG) in a~FS~CMa star, IRAS~17449+2320, was recently made by \mbox{\cite{Korcakova2022}}. This value of magnetic field modulus is comparable to the ones measured for the strongest Ap~stars. The space velocity component of this object, directed to the north pole of the Milky Way, is also unusually high. The most likely explanation for its nature is therefore a~post-merger.

The existence of an~FS CMa star with such a~strong magnetic field opens a~new window of opportunity for detecting post-mergers among B-type stars. This is what we decided to investigate, with the help of numerical simulations of open clusters in a~galaxy. The literature on stellar mergers is very sparse when it comes to lower-mass objects; however, as we intend to show in the following chapters, it is right in the domain of intermediate-mass stars (mainly B-type) where most merger events occur.

In the following sections, we explain the numerical methods used to calculate the aforementioned simulations as well as their analysis (Sects. 2 and 3). In Sect. 4, we bring forth the comparison of the results with a~sample of 32 FS~CMa stars, for which we were able to measure the system radial velocity, and therefore to calculate their space velocity components. In Sect. 5, a discussion of the results is provided, as are their implications.


\section{N-body simulations}
The present models were performed by the code \textsc{nbody6} \citep{Aarseth1999, Aarseth2003}.
The code combines the Ahmad-Cohen method \citep{Ahmad1973} with a fourth-order Hermite integrator \citep{Makino1991,Makino1992}.
Close interactions among stars were treated by regularising methods \citep{Kustaanheimo1965,Aarseth1974b,Mikkola1990}.
Stellar evolution of single stars as well as binaries (including mass transfer, common envelope (CE) evolution, and mergers) was dealt with using the algorithms of \citet{Hurley2000} and \citet{Hurley2002} for the solar metallicity of $Z = 0.014$.
This allowed us to follow the evolutionary stages of individual stars.
Details of the numerical algorithms employed by the \textsc{nbody6} code can be found in \citet{Aarseth1999,Aarseth2003}. It is important to emphasise the need to apply realistic conditions in order to make comparisons with observational data.


The present numerical models assume that all stars form in gas-embedded star clusters. 
The mass of the forming clusters is distributed 
according to the initial embedded cluster mass function (ECMF) of the form

\begin{equation}
\xi(M_{\rm ecl}) \equiv \frac{\text{d}N_{\rm ecl} (M_{\rm ecl})}{\text{d}M_{\rm ecl}} \propto M_{\rm ecl}^{-2},
\label{eecmf}
\end{equation}
where $\text{d}N_{\rm ecl}(M_{\rm ecl})$ is the infinitesimal number of embedded clusters with stellar masses in the range of $M_{\rm ecl}$ to $M_{\rm ecl} + \text{d}M_{\rm ecl}$ \citep[e.g.][]{Whitmore1999,Lada2003,Bik2003,FuenteMarcos2004,Gieles2006}. 
The ECMF of this form contains the same stellar mass within a~given logarithmic mass bin. 
Accordingly, we performed two realisations (i.e. the clusters have the same statistical properties but the masses of individual stars were resampled, and they were positioned differently in the phase space) of the most massive clusters ($M_{\rm ecl}$ = 8000~$\text{M}_{\odot}$) in our model set (see Table~\ref{sim_info} for the number of realisations for clusters of each mass). The most massive clusters are followed by a sequence of clusters of mass 8000~$\text{M}_{\odot}/2^n$, which are realised $2 \times 2^n$ times, where $n$ is an integer number ranging from $1$ to $7$.
This means that the least massive clusters have a stellar mass of 62.5~$\text{M}_{\odot}$; we calculated that there are $256$ clusters of this mass. The mass range of the model star clusters covers most of the mass range of open star clusters currently forming in the Galactic disc with the exception of the most massive clusters (e.g. \citealt{Zwart2010,Negueruela2011}).

\begin{table*}
    \caption{Numbers of simulations for a given mass of the clusters}     
    \label{sim_info}    
    \centering                          
    \begin{tabular}{l | c | c | c | c | c | c | c | c }        
    \hline\hline                 
    M[$M\odot$]  $\approx$               & 8000 & 4100 & 2040 & 1030 & 510 & 255 & 130 & 62         \\
    \hline
    \# simulations                       & 2    & 4    & 8    & 16   & 32  & 64  & 128 & 256 \\
    \hline
    \# stars per cluster $\approx$       &13 680& 7070 & 3590 & 1860 & 980 & 510 & 260 & 140 \\
    \hline
    \end{tabular}
\end{table*}

The desired physical quantities concerning stellar mergers of an entire stellar population were then obtained by investigating the whole population of modelled clusters, which reflect the composition of the stars forming in a~star-forming event in clusters of a~wide range of initial cluster masses. The embedded clusters initially consisted of two components: stars and gas. The stellar component is represented by the Plummer model \citep{Plummer1911} of mass, $M_{\rm ecl}$, and half-mass radius, $r_{\rm h}$, depending on the cluster mass according to eq. 7 of \citet{Marks2012}, which for the cluster masses in question ranges from $0.17$~$\mathrm{pc}$ to $0.31$~$\mathrm{pc}$. The clusters were initially in virial equilibrium and mass-segregated with the mass segregation parameter $S = 0.5$ (\citealt{Subr2008,Subr2012}). 
The initial positions and velocities of the stars were generated using the method of \citet{Aarseth1974b}. The gaseous component was modelled with the recipe of \citet{Kroupa2001b}; that is, the potential is an analytic Plummer sphere of a length-scale identical to the stellar component of the cluster, the initial total mass of the gas is two times larger than that of stars, and the gaseous mass stays unchanged for $0.6$~$\mathrm{Myr}$, whereupon it is exponentially reduced on a timescale of $r_{\rm h}$/(10 $\mathrm{km} \mathrm{s}^{-1})$ to approximate the feedback from young massive stars.

The clusters were subjected to the external gravitational field of the Galaxy, which was modelled by the three-component analytical model from \citet{Allen1991}. 
The clusters are on circular trajectories at a~Galactocentric radius of $8~\mathrm{kpc}$ and in the Galactic mid-plane.
This configuration enables us to study the stars not only within the clusters, but also in the tidal tails of the clusters \citep[e.g.][]{Kupper2008,Dinnbier2020b} as well as remote energetic escapers, which can move up to distances of several tens of $\mathrm{kpc}$ from their birth clusters.

Individual masses of the stars were drawn from the canonical \citet{Kroupa2001a} initial mass function. The minimum stellar mass is 0.08~$\text{M}_{\odot}$. The maximum allowed mass for a star formed in the cluster is a function of the cluster mass \citep{Weidner2010,Yan2023}. 
We assumed that all stars form as binaries. This assumption deals with issues of the angular momentum problem in star formation and was also used in the study by \citep{Kroupa1995a,Kroupa1995b}.
We ignored primordial triples or higher-order hierarchical systems because of uncertainty in the initial conditions.
However, if such systems later formed dynamically, they were treated self-consistently with the regularising methods of \cite{Aarseth1974CeMec}.
We treated differently binaries with masses below $m_{\rm thr} = 5$~$\text{M}_{\odot}$ and above this limit, which constitutes a threshold period mass.
The distribution functions of the mass ratio, orbital period, and eccentricity of binaries with a primary mass below $m_{\rm thr}$ were sampled from the birth distribution of \citet{Kroupa1995a} and subsequently subjected to pre-main-sequence eigenevolution \citep{Kroupa1995b}, which approximates the binary pre-main-sequence evolution. The distributions for orbital parameters of binaries with a primary mass above $m_{\rm thr}$ were taken from \citet{Belloni2017}, and are based on the observations of \cite{Sana2012} and \cite{Kobulnicky2014}. Such models were studied in the context of stellar ejection and runaway OB stars by \cite{Oh2015} and \cite{Oh2016}. Because of the short protostellar stage of massive stars, pre-main-sequence eigenevolution was not applied to stars of a mass above $m_{\rm thr}$. 
The adopted orbital period distribution results in the median orbital period of stars of a mass below $m_{\rm thr}$ initially being $\log_{10} (P/[\rm{days}]) = 5.5$, 
and the  median orbital period of stars of mass above $m_{\rm thr}$ initially being $\log_{10} (P/[\rm{days}]) = 2.0$. 
To test how the 5 $\text{M}_{\odot}$ threshold affects these results, we performed an equivalent 
set of simulations with the threshold period mass at 2 $\text{M}_{\odot}$. 
With 510 simulations in each of the two sets, we can approach a realistic distribution of open clusters in the Galaxy.

\section{Results}
Our aim is to analyse all merger events detected during the course of evolution of the simulations. Our focus is on the impact of mergers on the number of B-type stars. In the following, we shall answer questions regarding the spectral distribution of mergers, their dynamics, and stellar evolution. We discuss how many percent of mergers occur per individual spectral type and what is the mass distribution of the merger products as well as how many mergers occur before or after the CE phase and what is their final evolutionary stage.

\subsection{Distribution of spectral types}
The mass ranges for distinguishing individual spectral types are based on empirical masses for main-sequence stars from the work of \citet[Table VIII, p.232]{Habets1981}. The specific choices of mass boundaries for individual spectral types are shown in Table \ref{mb}. 
First, we investigated the spectral distribution among all the stars in the clusters at a time of 0 Myr in both sets of simulations. We found that around 70 \% of the stars are M-type stars. These stars have masses in the interval of 0.08 and 0.45~$\text{M}_{\odot}$. The second most common spectral type is the K type, lying slightly above 15~\% of the initial stellar population. Fewer than 4~\% of the stars are B-type stars.
The most common combination of spectral types in binaries at 0~Myr is the M-M binary.
Secondly, we looked at stars involved in merger events. After that, we were interested in the spectral types among the merger products themselves, and finally we calculated the fraction of mergers per individual spectral type.

\begin{table}[H]
\caption{Mass intervals for the spectral type analysis.}
\label{mb}
\centering                          
\begin{tabular}{c | c  }        
\hline\hline
Sp. type & M [$\text{M}_{\odot}$] \\
\hline
O    & $> 16$ \\
B    & 2.1 - 16 \\
A    & 1.4 - 2.1 \\
F    & 1.04 - 1.4 \\
G    & 0.8 - 1.04 \\
K    & 0.45 - 0.8 \\
M    & 0.08 - 0.45 \\
BrDw & $< 0.08$ \\
\hline
\end{tabular}
\tablefoot{Values are based on the empirical masses for main-sequence stars \citep{Habets1981}. Any star with a mass below 0.08 $\text{M}_{\odot}$ is considered a brown dwarf (BrDw).}
\end{table}

\begin{table*}[h!]
\caption{Distribution of individual spectral types for all stars at t~=~0~Myr, for stars involved in merger events and the final merger products.}
\label{percen}
\centering                          
\begin{tabular}{l | c | c | c | c | c | c | c | c}        
\hline\hline
Spectral&\multicolumn{2}{c|}{Stars}         & \multicolumn{2}{c|}{Stars involved}  & \multicolumn{2}{c|}{Merger}         & \multicolumn{2}{c}{Merger}\\
type    &\multicolumn{2}{c|}{at 0 Myr [\%]} & \multicolumn{2}{c|}{in mergers [\%]} & \multicolumn{2}{c|}{products [\%]}  & \multicolumn{2}{c}{ratio [\%]}\\
\hline
$m_{\rm thr}$ &2 $\text{M}_{\odot}$ & 5 $\text{M}_{\odot}$&2 $\text{M}_{\odot}$ & 5 $\text{M}_{\odot}$&2 $\text{M}_{\odot}$ & 5 $\text{M}_{\odot}$ &2 $\text{M}_{\odot}$ & 5 $\text{M}_{\odot}$   \\
\hline\hline
     &&&&&&\\
O        &      0.11 & 0.12        &    2.23 &  3.32 &  2.47 &  3.46 & 30.99 & 26.22  \\
\textbf{B}       &      \textbf{3.37} & \textbf{3.53}      &    \textbf{50.11}& \textbf{48.15}& \textbf{54.44}& \textbf{50.48}& \textbf{23.24} & \textbf{12.54} \\
A        &      2.64 & 2.77        &    13.42&  11.49&  14.28&  15.18& 7.77 & 4.80 \\
F        &      3.21 & 3.28        &    8.33 &  8.95 &  9.56 &  7.11 & 4.27 & 1.90 \\
G        &      3.95 & 4.05        &    6.75 &  6.53 &  5.71 &  3.88 & 2.07 & 0.84 \\
K        &      15.18& 15.29   &        9.94 &  10.71&  7.39 &  10.29& 0.70 & 0.59 \\
M        &      71.53& 70.97   &        9.22 &  10.82&  6.10 &  8.81 & 0.12 & 0.11 \\
BrDw &  0    & 0           &    0    &  0.02 &  0.06 &  0.78 & 0    & 0    \\
\hline
\end{tabular}
\tablefoot{The merger ratio shows what percentage of mergers is produced per spectral type. All values are shown for both threshold period masses, $m_{\rm thr}$:~2~$\text{M}_{\odot}$ and 5~$\text{M}_{\odot}$. In the case of B-type stars and $m_{\rm thr}$~=~5~$\text{M}_{\odot}$, the first column shows that 3.53 \% of all stars at 0 Myr were B-type stars. The second column shows stars involved in merger events (pre-mergers) and out of those there are 48.15 \% B-type stars. 50.48 \% of the merger products are of spectral type B. Finally, the last column shows that 12.54 \% of all B-type stars (calculated from the number of B-type stars at a time of 0 Myr) end up as merger products.}
\end{table*}

\begin{figure*}
\centering
\includegraphics[width=18cm]{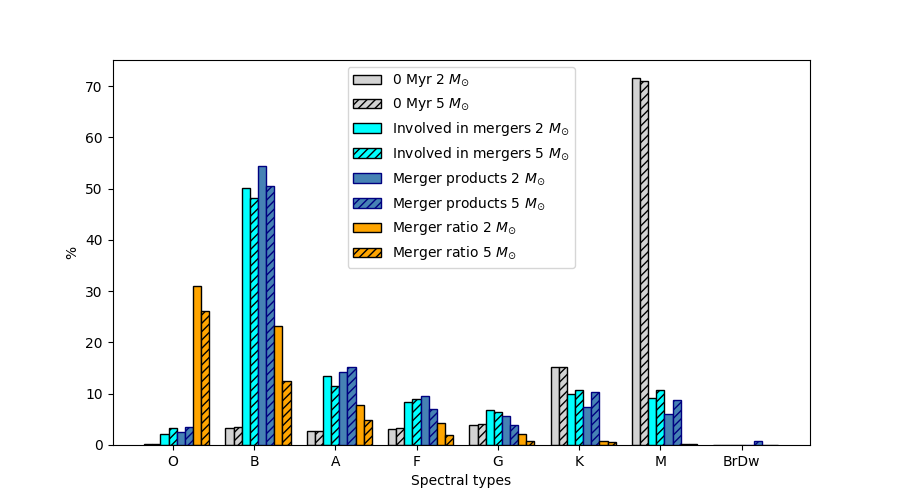}
  \caption{Bar chart representation of values in Table~\ref{percen}. The solid boxes represent the analysis of simulations with a threshold mass of 2~$\text{M}_{\odot}$. The hatched boxes represent simulations with a threshold mass of 5~$\text{M}_{\odot}$. The grey bars demonstrate all stars at a time of 0~Myr, the cyan colour stands for the initial spectral type distribution of stars that are involved in merger events, blue represents spectral types of the merger products, and the orange colour shows the ratio of mergers for each spectral type.
          }
     \label{percen_barchart}
\end{figure*}

In each set of simulations, we calculated the total merger ratio; that is, the ratio of the total number of mergers to the total number of all stars at t = 0 Myr in the given set of simulations.
The resulting percentage of mergers is 1.44 \% and 0.88 \% for $\text{m}_{thr}$ = 2 $\text{M}_{\odot}$ and $\text{m}_{thr}$ = 5 $\text{M}_{\odot}$, respectively.
The results of the analysis of individual spectral types for the two sets of simulations are shown in Table \ref{percen} (for a visual representation of Table \ref{percen}, see Fig. \ref{percen_barchart}. Included in Fig. \ref{percen_barchart} is a percentage for each spectral type showing how many mergers we detect per individual spectral type).
We can see that even if there are fewer than 4~\% of B-type stars at the beginning of the simulations, almost 50~\% of all stars involved in mergers (pre-mergers) are B-type stars. As a consequence of this, over 50~\% of the merger products are also of the B spectral type. The second most common spectral type among stars involved in merger events and among the merger products is the A type. 

\begin{figure*}
\subfloat{\includegraphics[width = 3.5in]{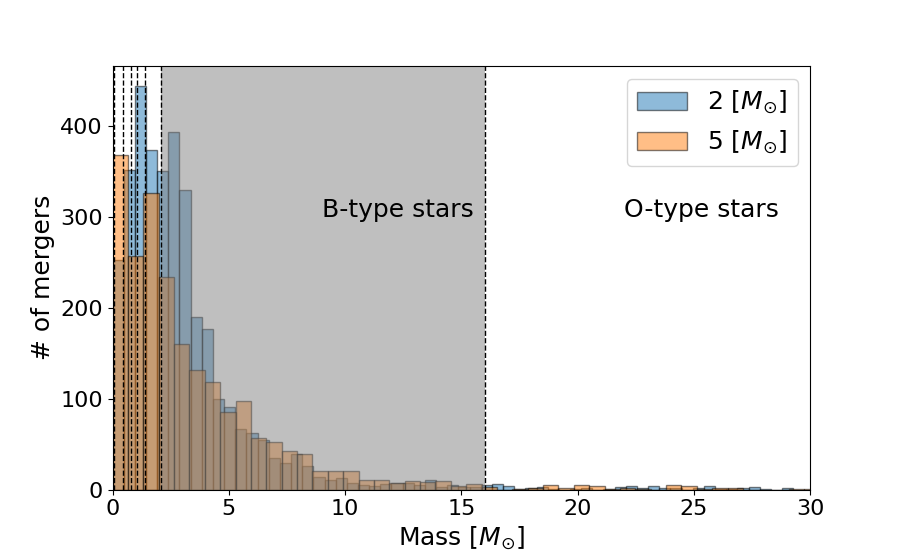}} 
\subfloat{\includegraphics[width = 3.5in]{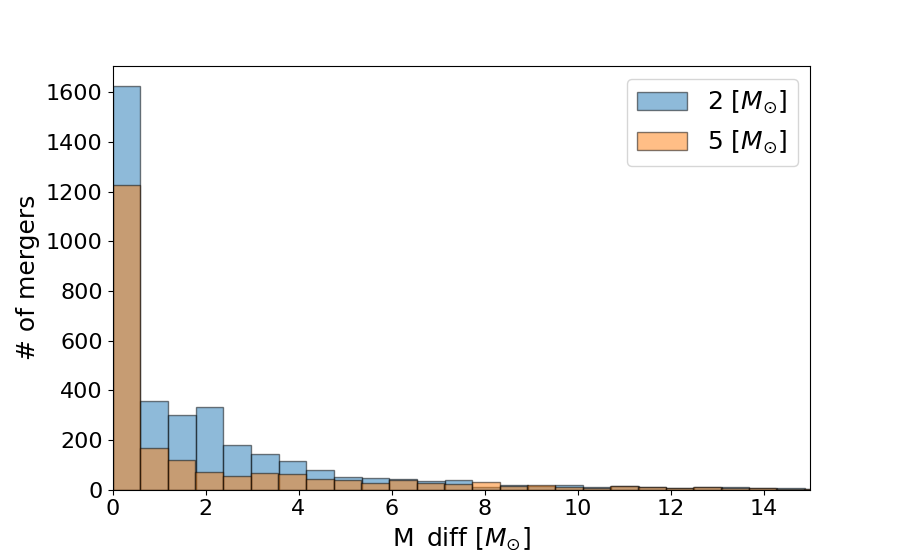}}
\caption{\textit{Left panel:} Mass distribution of the merger products (the spectral type B is marked by the grey region). 
\textit{Right panel:}~Distribution of mass differences of the primary and the secondary.}
\label{resultsmass}
\end{figure*}

In Fig. \ref{resultsmass}, we investigate the properties of the merger products. The left panel shows their mass distribution. We have highlighted the objects that we detect with masses between 2.1~$\text{M}_{\odot}$ and 16~$\text{M}_{\odot}$. We can see a peak in the number of mergers in the region of A-type stars, which corresponds to the observations of Ap~stars.  Their merger origin has already been suggested by \cite{Ferrario2009}.

In the right panel of Fig. \ref{resultsmass}, we plot a~histogram of mass differences between the primary and secondary of the preceding binary. The histogram shows that binaries with components similar in mass merge more often. There are, however, a~few cases with an extreme difference in primary and secondary masses (~>~30~$\text{M}_{\odot}$).

\subsection{Merger dynamics}

\begin{figure*}
\subfloat{\includegraphics[width = 3.5in]{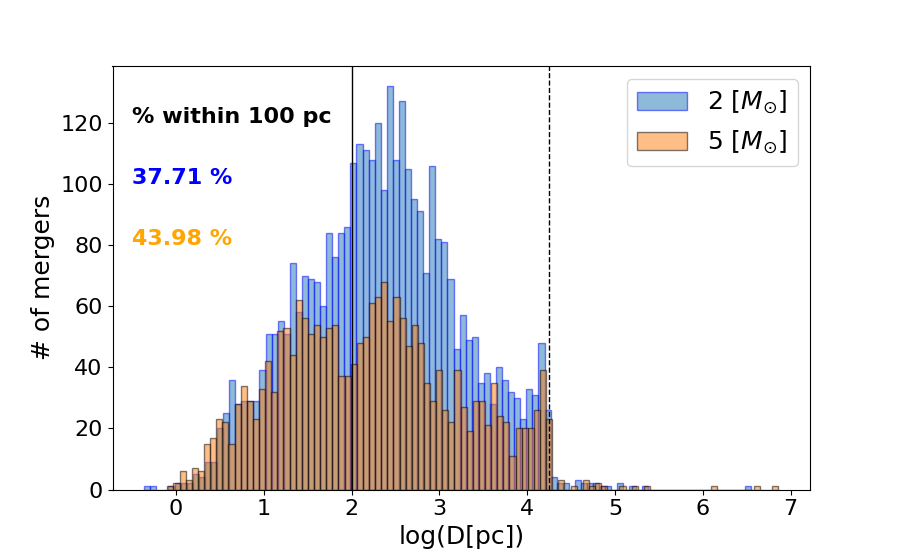}} 
\subfloat{\includegraphics[width = 3.5in]{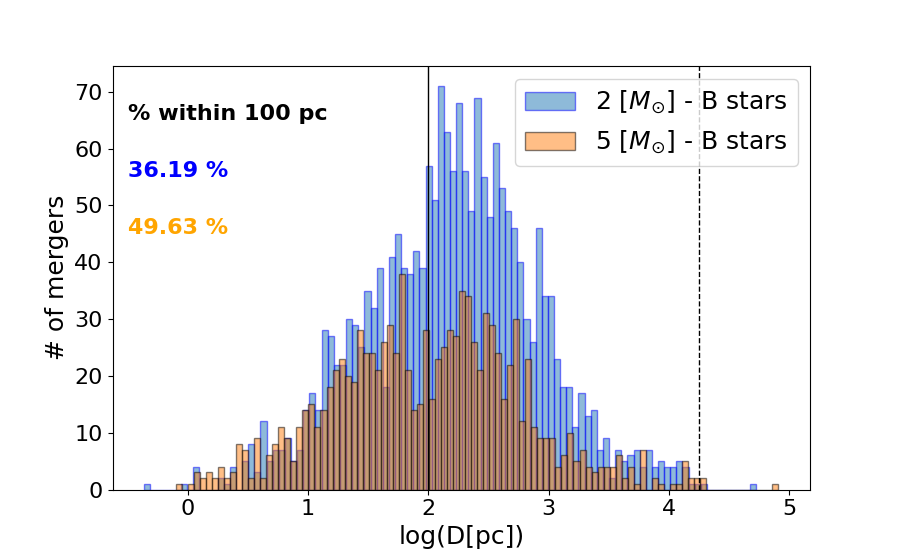}}
\caption{\textit{Left panel:} Histogram of the distance at the time of merger for the two sets of simulations of all spectral types. The results for a threshold period mass of 2~$\text{M}_{\odot}$ and 5~$\text{M}_{\odot}$ are depicted in blue and orange, respectively. The black line indicates a 100~pc distance. Percentages show the fraction of mergers that occurred within a 100~pc distance. The dashed black line highlights the cut-off distance at approximately 4.25 on the logarithmic scale. \textit{Right panel:} Distance histogram for B-type stars only. Colour coding corresponds to that of the left panel.}
\label{distances}
\end{figure*}

One of the interests of our research was the distance distribution of the merger products. A comparison of the two threshold period masses is given in the left panel of Fig.~\ref{distances}, which depicts the distances of the merger products right after they are produced. The distance was calculated from the centre of the respective birth cluster. For the clusters, which already dissolved, the distance is measured relative to the mass centre of the remaining stars. We can see for the 5~$\text{M}_{\odot}$ threshold period mass that the distribution shows a~bimodal profile, which could indicate that there is a~characteristic velocity of ejection of the pre-merger binary from the cluster with a~characteristic binary period that together lead to the ejected binary at preferred distances of $10^{1.5}$~pc and $10^{2.3}$~pc from the cluster. Alternatively, the bimodal profile could be caused by spatial clustering of escaping stars near the cusps of their epicycles, the point of which is expected to be at $10^{2.2}$~pc.

\begin{figure*}

\centering
\subfloat{\includegraphics[width=9cm]{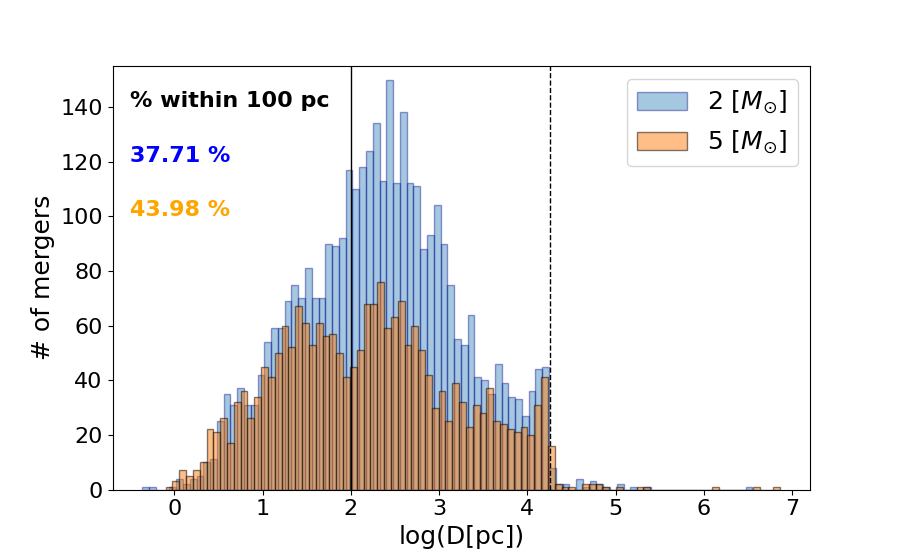}}\hfil
\subfloat{\includegraphics[width=9cm]{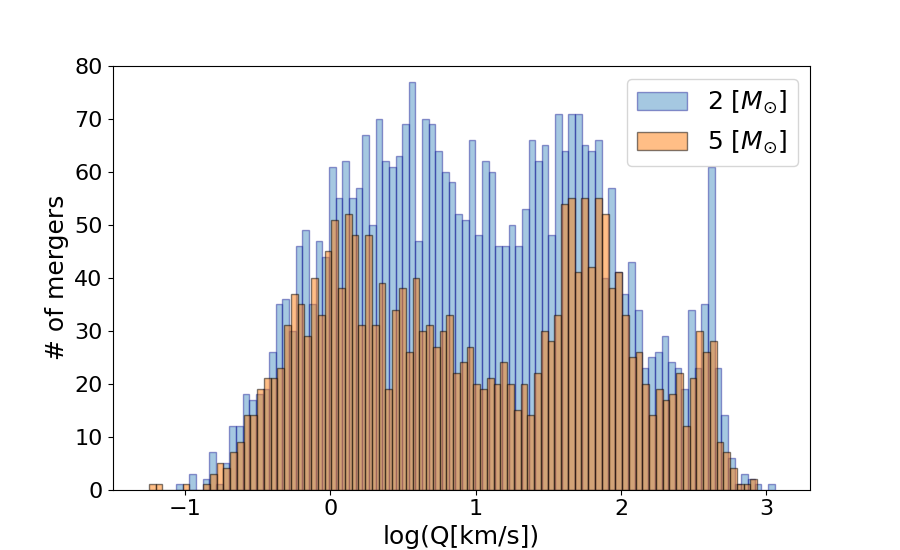}}

\subfloat{\includegraphics[width=9cm]{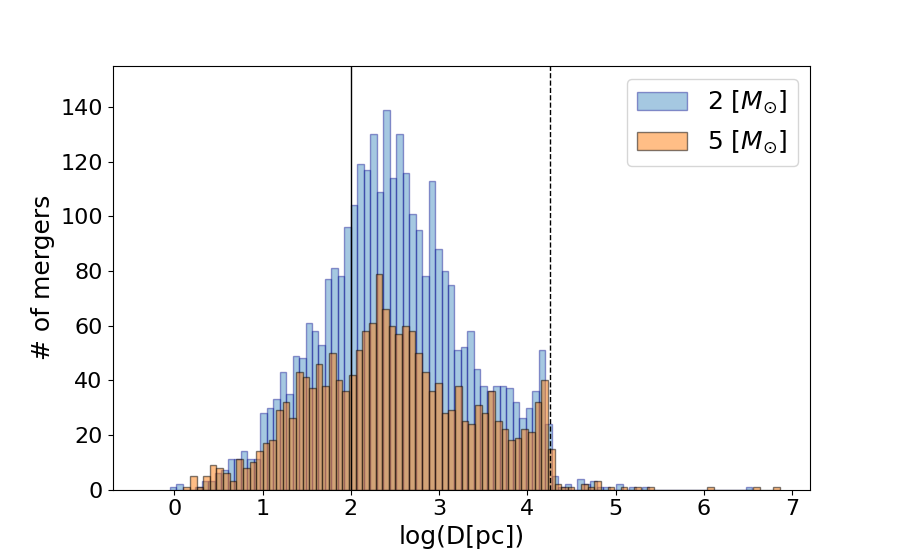}}\hfil
\subfloat{\includegraphics[width=9cm]{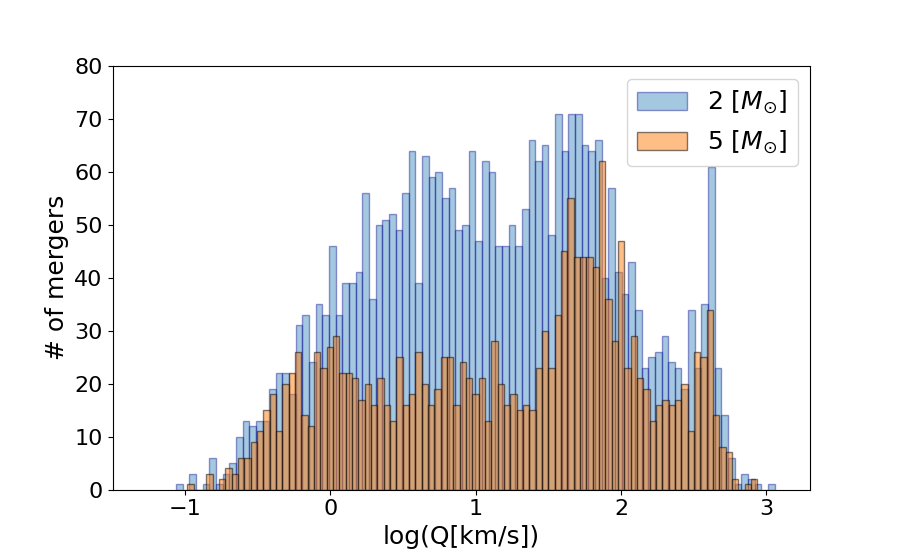}}

\subfloat{\includegraphics[width=9cm]{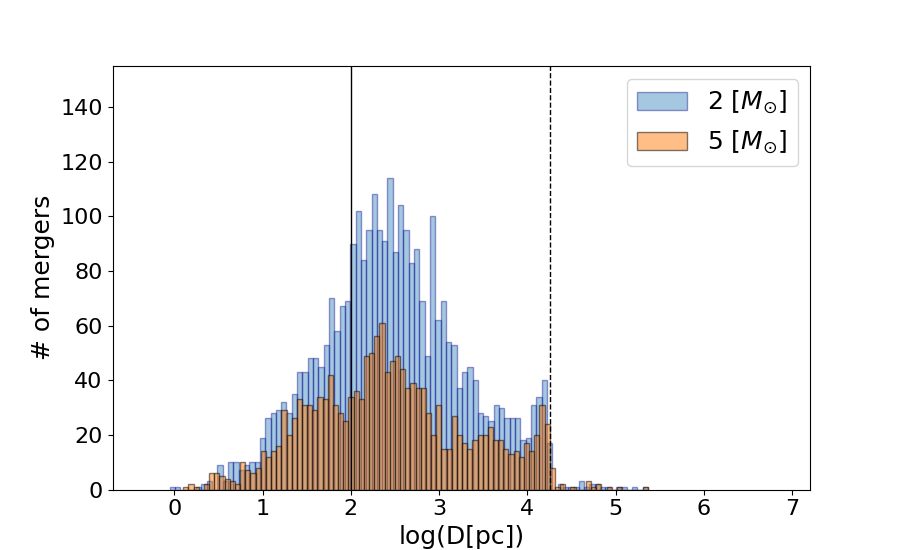}}\hfil
\subfloat{\includegraphics[width=9cm]{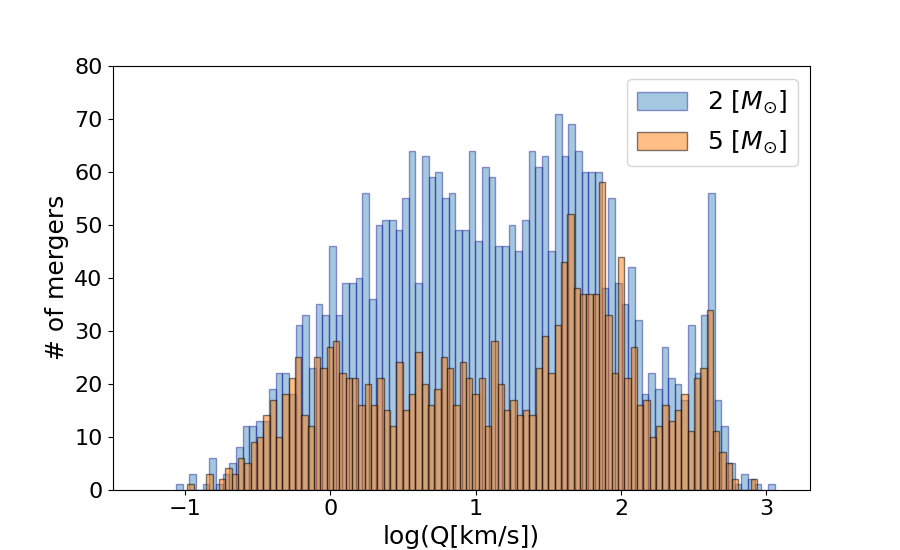}}
\caption{Effect of the pre-main-sequence mergers and main-sequence mergers on the total distribution of mergers. \textit{Left column:} Histogram of the distances of mergers at the time of the merger event. \textit{Right column:} Space velocity, $Q$, of mergers, relative to the centre of mass of the cluster, at the time of the merger event. The upper panels show the distribution for all mergers, and the middle panels the distribution in which the contribution of the mergers during the pre-main-sequence eigenevolution was removed. The properties of mergers created after the main-sequence are shown in the bottom panels (no early mergers and no mergers before CE).} The notation is the same as in Fig. 3.
\label{hist_D_Q}
\end{figure*}

The same result is seen if we look at the distances of mergers of the B-type stars only (see the right panel of Fig.~\ref{distances}). The dip in the distance profile is placed at a distance of about 100~pc from the centre of the cluster. The simulations with the 2~$\text{M}_{\odot}$ threshold period mass produced 3558 mergers (compared with 2167 mergers produced by the 5~$\text{M}_{\odot}$ set of simulations) and the bimodal profile is not as visible there. 
In order to better understand this bimodal distribution, we plot distance histograms in Fig. \ref{hist_D_Q}, in which we compare the original distance histogram from the left panel of Fig. \ref{distances} of all mergers with the distance histogram without the early mergers and the histogram without mergers that occurred before the CE phase (see histograms on the left in Fig. \ref{hist_D_Q}). The right panel of Fig.~\ref{hist_D_Q} shows the space velocity of the mergers relative to the centre of mass of the cluster (for more details, see Sect. 4). We can see that the pre-main-sequence mergers significantly contribute to the first peak of the distance distribution and this effect is even stronger in the velocity distribution. From an observational point of view, most FS~CMa stars' locations cannot be traced to their original clusters, with the exception of two cases observed by \cite{Fuente2015} in which the FS~CMa star is located at the centre of the young stellar cluster.

Both sets of simulations seem to show a cut-off at around 17.7~kpc (log(D[pc])~=~4.25), which corresponds to mergers occurring throughout the galactic disc, but we detected several cases of mergers beyond this limit. In the case of $m_{\rm thr}$~=~2~$\text{M}_{\odot}$, we found 27 merger events happening beyond the cut-off distance and one of these ends up as a massless remnant. Simulations with $m_{\rm thr}$~=~5~$\text{M}_{\odot}$ yielded 25 mergers beyond the cut-off distance, and four result in massless remnants. \cite{Monari2018} measured the Galactic escape speed curve and derived an escape velocity at the Galactocentric distance of the Sun of 580~$\pm$~63~km/s. We found 14 cases of binaries with a velocity over this value at the time they merged in both sets. Seventeen more mergers have velocities within the margin of error. In four cases, the merger occurred at a distance larger than the cut-off distance.

\subsection{Stellar evolution}
\begin{figure*}
\subfloat{\includegraphics[width = 3.5in]{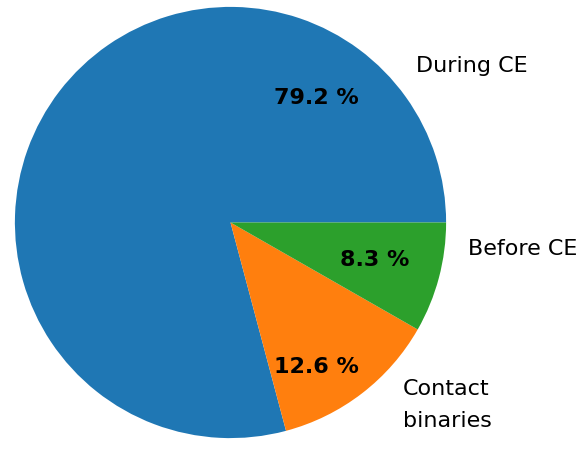}} 
\subfloat{\includegraphics[width = 3.5in]{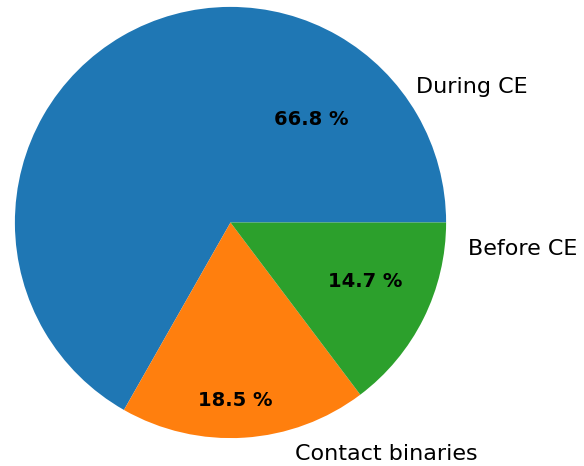}}
\caption{Pie charts showing the percentages of mergers that occurred during or before the CE phase or those that come from contact binaries. \textit{Left panel:} Simulations with $m_{\rm thr}$ = 2 $\text{M}_{\odot}$. \textit{Right panel:} Simulations with $m_{\rm thr}$ = 5 $\text{M}_{\odot}$}.
\label{pie}
\end{figure*}

The N-body code allows us to tell whether a merger happened before or during the CE phase of binary evolution and how many mergers come from contact binaries. 
All three cases for the threshold period mass of 2~$\text{M}_{\odot}$ and 5~$\text{M}_{\odot}$ are visualised in Fig.~\ref{pie}. In both simulation sets, most mergers went through the CE phase: 79.2~\% in the case of $m_{\rm thr}$ 2~$\text{M}_{\odot}$ and 66.8~\% in the 5~$\text{M}_{\odot}$ case. Mergers that come from contact binaries account for 12.6~\% and 18.5~\% of cases for $m_{\rm thr}$ of 2~$\text{M}_{\odot}$ and 5~$\text{M}_{\odot}$, respectively. Notably, 8.3~\% and 14.7~\% of mergers for 2~$\text{M}_{\odot}$ and 5~$\text{M}_{\odot}$, respectively, occurred before they entered the CE phase. A possible explanation for such cases is an interaction with a~third star resulting in a~binary that is kicked out of the cluster with high speed. This binary usually has a~short semi-major axis and a~large eccentricity. Due to the effect of tidal forces, the orbit of this binary will circularise, causing the semi-major axis to shrink, which may lead to a~merger. The merger event may also be supported by the expansion of the stellar radius during the stellar evolution on the main sequence.

\begin{figure}
\centering
\includegraphics[width=9.5cm]{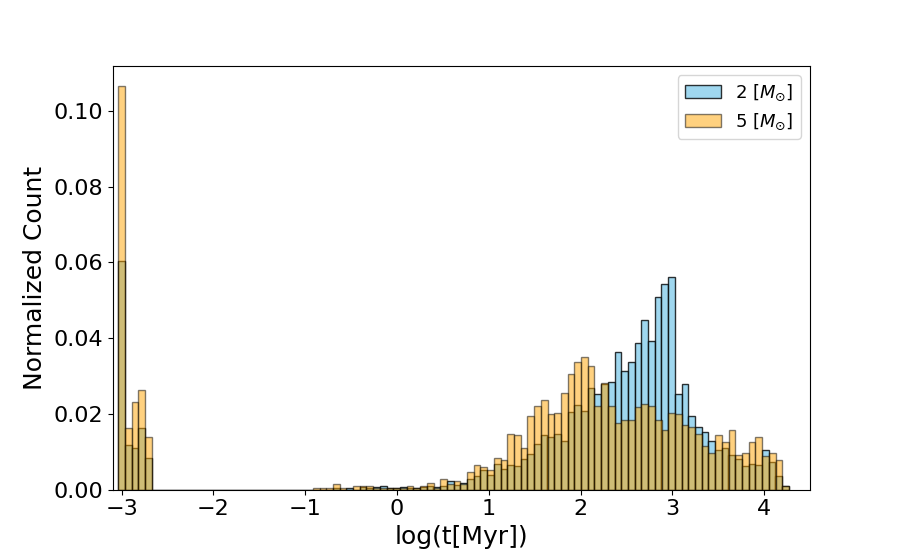}
  \caption{Histogram of the time of occurrence of individual merger events. We can clearly see the early mergers resulting from the pre-main-sequence eigenevolution process. All these early mergers occurred within the first 2000 years of the simulations. The second broad maximum in the histogram represents all the merger events that happened after the first 100 Myr.}
     \label{t_hist}
\end{figure}

\begin{figure} 
\begin{minipage}[b]{1.0\linewidth}
    \centering
    \includegraphics[width=9cm]{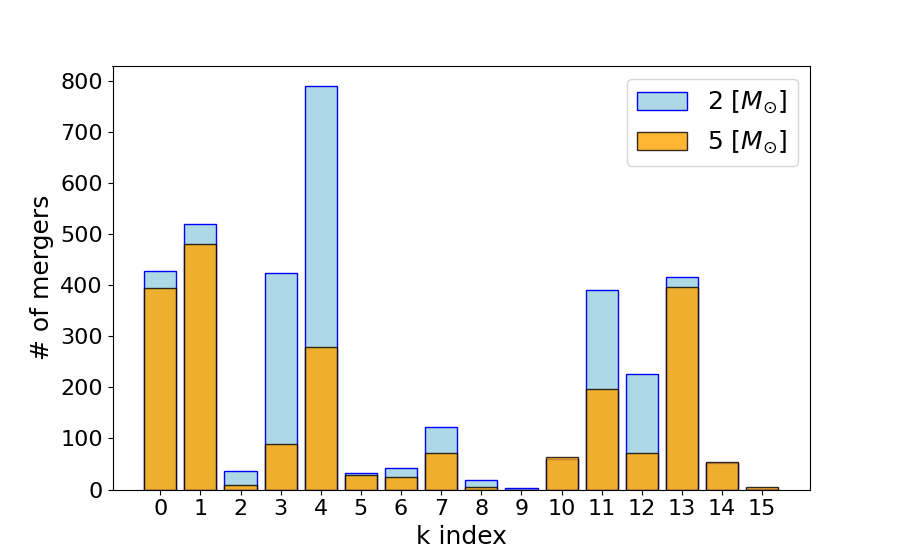}
    \caption{Bar chart of the k~index of the merger products. The entire ensemble of individual stars accounts for 3558 mergers for the simulations with $m_{\rm thr}$ = 2~$\text{M}_{\odot}$ (in blue) and 2167 mergers in case of the simulations with $m_{\rm thr}$ = 5~$\text{M}_{\odot}$ (in orange). An explanation of the k~index values can be found in Table~\ref{kindx}. 
      }
    \label{k_hist}
\end{minipage}\hfill
\end{figure}

\begin{figure*}
\subfloat{\includegraphics[width = 3.5in]{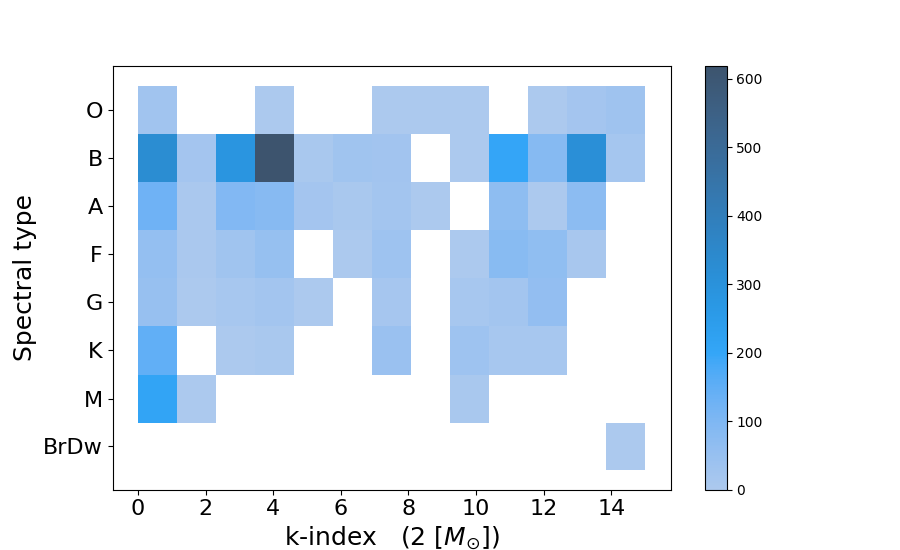}} 
\subfloat{\includegraphics[width = 3.5in]{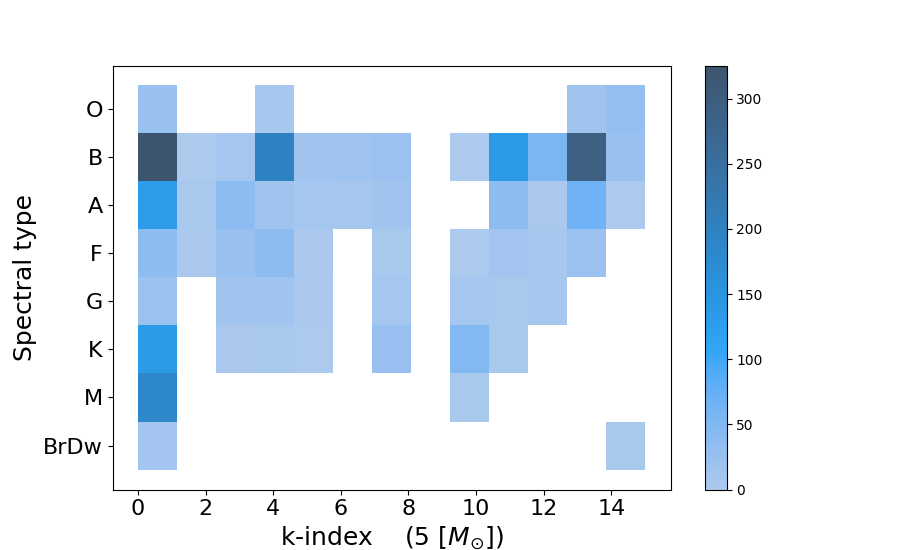}}
\caption{3D histograms \textbf{of} the k-index distribution for individual spectral types. \textit{Left panel:} Results of the simulations with the 2 $\text{M}_{\odot}$ threshold period mass. \textit{Left panel:} Results for the 5 $\text{M}_{\odot}$ threshold period mass.}
\label{k_spt}
\end{figure*}

The evolutionary stage of the merger product is shown in Fig.~\ref{k_hist}. It is represented by the evolutionary index, k, which is explained in Table~\ref{kindx}. We can see that a~lot of merger products end up on the main sequence. A~significant number of events lead to k~=~4 and we detect over 20~\% of compact objects such as neutron stars and black holes in simulations with $m_{\rm thr}$~=~5~$\text{M}_{\odot}$. These also include a~significant number of white dwarfs.
To demonstrate what the distribution of evolutionary stages looks like in more detail, we plot a~3D histogram (see Fig.~\ref{k_spt}) in which individual spectral types are resolved. Given the number of B-type merger products, it is not surprising that this spectral type also dominates in certain evolutionary stages.

The initial conditions for the N-body code follow the result of pre-main-sequence eigenevolution (as is described in Sect. 2). The reducing of the semi-major axis during the pre-main-sequence eigenevolution leads in some cases to the binary components being so close to each other that they merge immediately upon the start of the N-body simulations (or within the first 2000 years). This is a non-negligible number of mergers, which is included in our sample and constitutes $\approx$~10~\% and $\approx$~18~\% of the sample in simulations with $m_{\rm thr}$ of 2~$\text{M}_{\odot}$ and 5~$\text{M}_{\odot}$, respectively. They can be seen in Fig.~\ref{t_hist}, which shows a logarithmic scale of the times of the merger events. In order to include mergers that occurred at t = 0 Myr in the logarithmic scale, we moved them to t~=~1000 years. The histogram is normalised to the total number of mergers. We register more of the early merger events in the case of simulations with $m_{\rm thr}$ of 5~$\text{M}_{\odot}$. Conversely, in the case of merger events occurring from 100 Myr onwards, simulations with $m_{\rm thr}$ of 2~$\text{M}_{\odot}$ produce almost twice as many mergers than the simulations with $m_{\rm thr}$~=~5~$\text{M}_{\odot}$. The main reason for the significantly higher number of mergers in the case of 2~$\text{M}_{\odot}$ simulations is a higher number of contact binaries present in these clusters. This naturally leads to more mergers. The higher registered number of the early mergers in case of the $m_{\rm thr}$~=~5~$\text{M}_{\odot}$ simulations is because the pre-main-sequence eigenevolution is included only for the low-mass binaries (those for which the primary mass does not exceed the $m_{\rm thr}$), which means there are more stellar binaries undergoing pre-main-sequence eigenevolution than in the models with $m_{\rm thr}$~=~2~$\text{M}_{\odot}$.

\begin{table}[h!] 
\begin{minipage}[b]{1.0\linewidth}
    \centering
    \caption{K index of the evolutionary stages.}
    \label{kindx}
    \begin{tabular}{ l || c }
    \hline
    \textbf{0} & MS fully convective  \\
    \hline
    \textbf{1} & MS \\
    \hline
    \textbf{2} & Hertzsprung Gap \\
    \hline
    \textbf{3} &  First Giant Branch\\
    \hline
    \textbf{4} & Core Helium-Burning\\
    \hline
    \textbf{5} &  Early AGB\\
    \hline
    \textbf{6} &  Thermally Pulsing AGB\\
    \hline
    \textbf{7} &  Naked Helium Star MS \\
    \hline
    \textbf{8} & Naked Helium Star Hertzsprung Gap\\
    \hline
    \textbf{9} &  Naked Helium Star Giant Branch\\
    \hline
    \textbf{10} &  Helium White Dwarf\\
    \hline
    \textbf{11} &  Carbon/Oxygen White Dwarf\\
    \hline
    \textbf{12} &  Oxygen/Neon White Dwarf\\
    \hline
    \textbf{13} &  Neutron Star\\
    \hline
    \textbf{14} &  Black Hole\\
    \hline
    \textbf{15} &  massless remnant\\
    \hline
    \end{tabular}
    \tablefoot{K-index definition is taken from \cite{Hurley2000}.}
    \end{minipage}
\end{table}

\section{Comparison with observations}

We selected a~sample of FS~CMa stars from a~list of confirmed or candidate objects by Miroshnichenko.\footnote{The list of confirmed or candidate FS CMa objects can be found at \url{https://home.uncg.edu/~a_mirosh/hswd/main/FSCMa_objects_full_list_Sept2013.html}.} The availability of high-quality high-resolution spectra was our main selection criterion. For most of the known FS~CMa stars, it is not possible to trace affiliation to their parent stellar cluster, so the distances of the mergers we analysed in Sect. 3 therefore cannot be compared to observations. Instead, we calculated radial velocities for our sample in order to compare their space velocities. Most of the spectra were gathered from the archives of the European Southern Observatory (ESO) and the Canada-France-Hawaii Telescope (CFHT), with the exception of MWC~137 and AS~160 for which we also gained new high-resolution eschelle spectra from the UVES spectrograph on the VLT telescope UT2. We also collected data from FEROS (ESO), UVES (ESO), Xshooter (ESO), and ESPaDOnS spectrographs (CFHT). A brief summary of the accumulated data is given in Table~\ref{fscma_info}.

\begin{table}
\caption{Spectral resolving power and wavelength range of the spectrographs.}
\label{fscma_info}
\centering                          
\begin{tabular}{l | c | c}        
\hline\hline                 
Spectrograph & Spec. Res & $\lambda_{range} \text{[\AA]}$ \\    
\hline                        
FEROS           &       48 000  &       3 528 - 9 217   \\
UVES (blue)         &   80 000  &       3 732 - 5 000   \\
UVES (red)          &   110 000 &       5 655 - 9 464   \\
XSHOOTER            &   18 400  &       5 337 - 10 200 \\
ESPaDOnS        &       68 000  &       3 700 - 10 500 \\
UVES (new)      &       66 320  &       3 527 - 9 216   \\ 
\hline                                   
\end{tabular}
\end{table}

The radial velocity of each object was measured using the forbidden oxygen lines $\lambda\lambda$ [O~I]~6300.304 and 6363.776~\AA. FS~CMa stars are known to be variable spectroscopically as well as photometrically on many different timescales and these narrow emission lines can be stable on timescales of months to years~\citep{Kucerova2013}. In the region around the [O~I]~6300.304~\AA\ line, there are telluric lines that in some cases affect the line profile significantly. The [O~I]~6363.776~\AA\ line is, on the other hand, weaker. It is too weak in the case of IRAS~06071+2925 to properly fit the profile and the resulting radial velocity was only measured for the stronger oxygen line. For better precision, we took into account both oxygen lines wherever possible. From the list of FS~CMa stars (those with high-resolution spectra available), we only used the ones for which we were able to clearly identify the oxygen lines. After this selection, our sample contains 32 stars.

Using IRAF\footnote{IRAF is distributed by the National Optical Astronomy Observatories, operated by the Association of Universities for Research in Astronomy, Inc., under contract to the National Science Foundation of the United States.} (Image Reduction and Analysis Facility, \citealt{Iraf1986}) we checked each spectrum to see whether the heliocentric correction had been made. To fit the [O~I] lines, a~Gaussian profile available in IRAF was used to determine their central wavelengths. For most objects, the error in measurements of radial velocity was estimated from the average value of a~number of radial velocity measurements. In those cases in which we only had one spectrum available for the radial velocity determination, we estimated the error from a~fit of the oxygen lines.

Our measured radial velocities in combination with the \textit{Gaia} data provide space velocities relative to the local standard of rest, which we can directly compare with the results of our simulations. We used the values $\text{(U,V,W)}_{\odot}$ = (-8.5, 13.38, 6.49) km/s of \cite{Coskunoglu2011} to correct for the solar motion. The results of the radial velocity measurements and the calculated space velocity, Q, together with the U, V, and W components, followed by eq. \ref{Qobs}, can be found in Table~\ref{velocities}, including the space velocity components U, V, and W.

\begin{equation}
    Q_{obs} = \sqrt{U^2 + V^2 + W^2}
\label{Qobs}
.\end{equation}

The space velocities of the simulated mergers were calculated at the time of the merger and with respect to the centre of mass of the corresponding cluster. The velocities of the simulated mergers were calculated as follows:

\begin{equation}
    Q_{sim} = \sqrt{v_x^2 + v_y^2 + v_z^2},
\end{equation}

\noindent where $v_x$, $v_y$, and $v_z$ are the components of the velocity vector.
In order to be able to compare the simulated post-mergers with the observed sample of FS~CMa stars, we converted the velocities to the Galactocentric frame. To this end, we used the velocity vector of the individual clusters at the time of each mergers to calculate the Galactocentric velocity in our simulations. In the case of the observed stars, we assumed a circular velocity of 220 km/s for the Sun and the previously mentioned vector of solar motion (U,V,W)$_{\odot}$.

\begin{figure*}
\subfloat{\includegraphics[width = 3.8in]{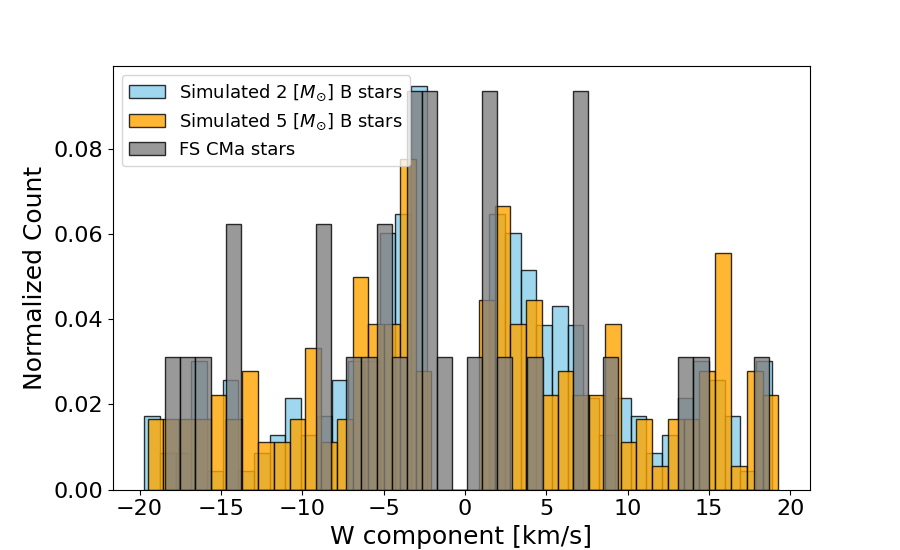}} 
\subfloat{\includegraphics[width = 3.6in]{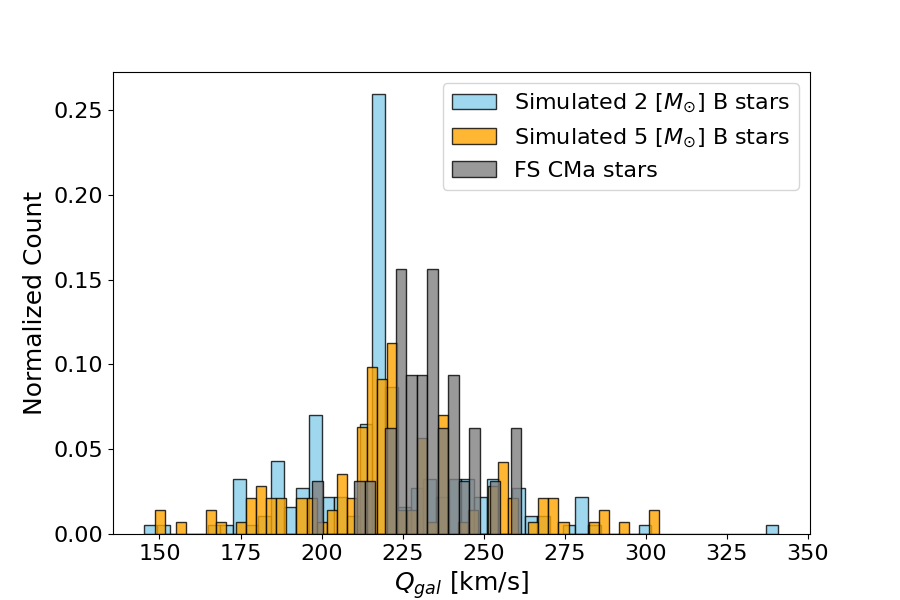}}
\caption{Comparison of the simulated mergers with observed FS~CMa stars. \textit{Left panel:} W component of the space velocity. Due to the low number of extremely fast FS CMa stars, the displayed range of velocities is restricted from -20 to -2~km/s and 2 to 20 km/s (the full distribution is shown in Fig. \ref{hist_Q_obs}). The region around a zero value for the W velocity has also been removed. Even if the majority of simulated mergers occupies this region, it is impossible to compare them with observations due to high extinction in the galactic plane. \textit{Right panel:} Same sample of mergers, plotted in space velocity with respect to the galactic centre.}
\label{w_cutoff}
\end{figure*}

In the left panel of Fig. \ref{w_cutoff}, we show a~normalised histogram of the W component of the space velocities for the simulated B-type mergers from both sets and for the observed FS~CMa stars. In order to see the comparison of the observed and simulated data clearly, we removed the region around the galactic plane in Fig.~\ref{w_cutoff}. The high number of mergers with the W velocity component around 0 km/s can be seen in Fig. \ref{hist_Q_obs}, in which we show only the positive values of a symmetric distribution. The observational bias is likely caused by the presence of dust in the disc of the Milky Way, which prevents us from observing any FS~CMa stars in its plane. We only show velocities from -20~km/s to -2~km/s and from 2~km/s to 20~km/s, corresponding to the ranges of observed velocities. The W component of the velocity vector of the FS~CMa stars from our sample do not go beyond these values. The W component of the space velocities of the simulated mergers reaches values from -400 km/s to around 1000 km/s.

\begin{figure*}
\subfloat{\includegraphics[width = 3.47in]{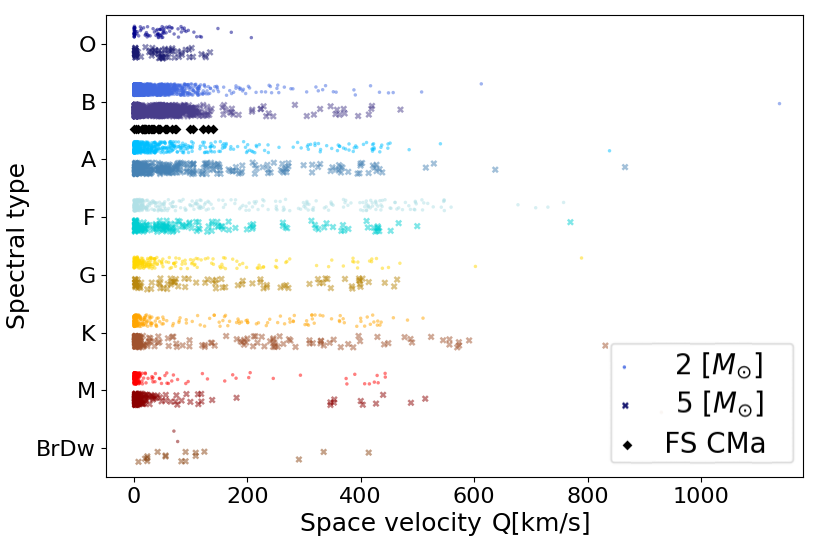}} 
\subfloat{\includegraphics[width = 3.5in]{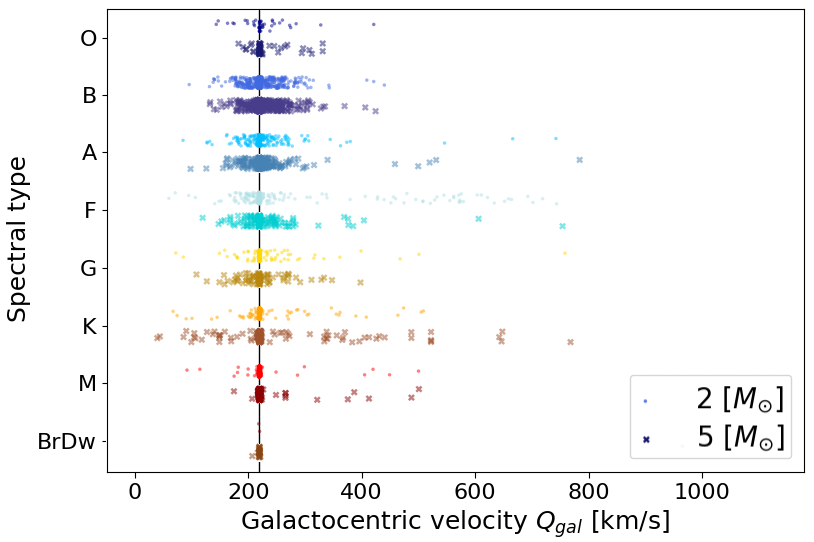}}
\caption{Comparison of space velocities of individual spectral types of mergers. The lighter coloured dots represent the simulations of a 2 $\text{M}_{\odot}$ threshold period mass and the darker coloured crosses represent the 5 $\text{M}_{\odot}$ simulations. \textit{Left panel:} Space velocities with respect to the centre of mass of the parent cluster. Measured space velocities of the FS CMa stars in our sample are included in the plot under spectral type B (black diamonds). \textit{Right panel:} Comparison of Galactocentric space velocities of the simulated mergers. The circular velocity of the corresponding cluster is denoted by the vertical line at a velocity of 219.7 km/s. We can see that the less massive the post-mergers are, the further they spread out from their original cluster.}
\label{SPTqsim_qgal}
\end{figure*}

A comparison of space velocities between the simulated mergers and the observed FS CMa stars is provided in Fig. \ref{SPTqsim_qgal}, in which we show individual spectral types. The space velocity with respect to the local standard of rest calculated from the measured radial velocities and the \textit{Gaia} data is shown in the left panel. The right panel of Fig. \ref{SPTqsim_qgal} shows the Galactocentric velocity of the simulated mergers. The position of the individual clusters of the simulations is denoted by the black line at the velocity of 219.7 km/s. As can be seen, the lower-mass post-mergers stray further from their original cluster. This trend is possible to see for both threshold period masses. Additional comparison of space velocities, Q$_{sim}$ (with respect to the centre of mass of the cluster), and Q$_{obs}$ (with respect to the local standard of rest) can be found in Fig. \ref{hist_D_Q}.
This histogram is normalised by the total number of data points. The histogram in Fig. \ref{hist_Q_obs2} also exhibits two distinct peaks in its distribution. The peak at larger distances is in good agreement with the distribution of the observed FS CMa. The peak closer to the cluster (at  Q$_{sim}$ = 0 km/s) corresponds in vast majority to mergers, whose W component of the space velocity is between -2 and 2 km/s. We suspect that more FS~CMa stars are hidden in the dusty disc of our Galaxy, where it is too difficult to observe them. Our sample contains only a small number of stars, but it is the FS CMa group that provides the best chance of finding more post-merger objects when only a fraction of real mergers is known.

\section{Discussion and conclusions}
B-type stars, especially those in the lower mass range, present an important class of objects in research into mergers and their effect on the interstellar environment. As we showed in our N-body simulations, around 50 \% of stars involved in merger events are B-type stars. Of all B-type stars, 23.24 \% (for $m_{\rm thr}$~=~2~$\text{M}_{\odot}$) and 12.54 \% (for $m_{\rm thr}$~=~5~$\text{M}_{\odot}$) are mergers (see Table \ref{percen}). \cite{Renzo2019} performed population synthesis calculations for isolated binaries and among their main results is that a fraction of $\approx$~22~\% of their binaries result in stellar mergers.

From our analysis, we can see that more than 50~\% of the resulting merger products are of spectral type B (see Fig. \ref{resultsmass} (a)). Our mass distribution histogram shows a higher number of merger products in the lower mass range of B-type stars. The peak in the amount of mergers is reached in the region of A-type stars and between 4.8~\% and 7.77~\% of all A-type stars are mergers (see Table \ref{percen}). These results are in agreement with the observations of Ap~stars \citep{Berdyugina2009}. We could find progenitors of massive magnetic Ap~stars among the late B-type FS~CMa post-mergers. One suggested scenario is that a~post-merger product is out of thermal equilibrium and the stars become overluminous for a period of time. At this stage, the stars could exhibit the properties of the FS~CMa stars. After they reach the equilibrium and settle back on the main sequence, we might in some cases observe them as A-type stars. \cite{Schneider2020} describe this evolution scenario in their work regarding O~stars.

Only around 3 \% of all merger products are O-type stars. The percentage of O-type stars at t=0 Myr is around 0.1 \% and the fraction of mergers among O-type stars is around 30 \%. It is important to note that although O-type stars typically appear not only in binaries, but also in multiple systems, these are not accounted for in the simulations.
Observational evidence shows that B stars have on average from $1.1$ companions (for 3 $\text{M}_{\odot}$ stars) to $1.9$ companions (for 16 $\text{M}_{\odot}$ stars) \citep{Moe2017,Offner2023}. A companion frequency of at least two is also implied from a comparison to companions to Cepheids, which evolve from mid-B stars, by \cite{Dinnbiersub}. 

Our results also show a~significant number of mergers that happened before entering the CE phase. We see 8.3~\% and 14.7~\% of all merger products for the threshold period mass of 2~$\text{M}_{\odot}$ and 5~$\text{M}_{\odot}$, respectively. This could be the result of an~interaction of a~binary with a~third star in the birth cluster. The outcome of such an interaction can result in a~binary that is kicked out of the cluster with a~high velocity. A~binary system like this will often be hardened, and thus have a~short semi-major axis and a~large eccentricity. As a~result of tidal forces, the orbit will circularise, while the semi-major axis is reduced and the binary may end up merging. Stellar evolution on the main sequence may cause these binaries to result in a merger as well.
12.6 \% and 18.5 \% (for the 2~$\text{M}_{\odot}$ and 5~$\text{M}_{\odot}$ simulations, respectively) are the cases of mergers of contact binaries. An observational example of a merger event of a contact binary can be found in \cite{Tylenda2011}, who analysed the photometry of V1309 Sco from the OGLE (Optical Gravitational Lensing Experiment) project. Outbursts of V1309 Sco are currently explained by a~stellar merger.

The final evolutionary stage represented in Fig.~\ref{k_hist} shows high numbers of merger products that end up on the main sequence or in the core helium-burning stage. In a~lot of cases, the final product is a~compact object, most commonly a~neutron star. Simulations with a threshold period mass of 2~$\text{M}_{\odot}$ produce more mergers in general; however, the most significant difference from the second set of simulations with $m_{\rm thr}$~=~5~$\text{M}_{\odot}$, in regard to their final evolutionary stage, can be seen in the number of objects on the first giant branch, the core helium-burning stars, Ca/O or O/Ne white dwarfs, where there is often more than double the number. The total number of mergers registered in the simulations with $m_{\rm thr}$ of 2~$\text{M}_{\odot}$ is almost twice as high as for the simulations with $m_{\rm thr}$ of 5~$\text{M}_{\odot}$. As can be seen in Fig \ref{t_hist}, the opposite is true for the case of early mergers (within the first 2000 years), which are the consequence of pre-main-sequence eigenevolution.
Due to a higher number of close binaries in simulations with $\text{m}_{thr}$~=~2~$\text{M}_{\odot}$, we detect more mergers.


We provide new measurements of system radial velocities for 32 FS~CMa stars. Accurate measurements of radial velocities or of any fundamental properties are very difficult, given the complexity of the spectra of these objects. In some cases, the signal-to-noise ratio is not very high or the spectral line is very weak, and both these factors can contribute to the uncertainties of our provided radial velocities. We have managed to provide new values for some of the stars in our sample. These results, in~combination with \textit{Gaia} data, were used to calculate the space velocities of the 32 FS~CMa stars. A~comparison of the W component of the space velocity vectors has been made for B-type mergers and the FS~CMa stars in the left panel of Fig.~\ref{w_cutoff}. Both simulation sets and the observed values are normalised by the total number of data points. 
The distribution of Galactocentric space velocities of mergers from the right panel of Fig. \ref{w_cutoff} agrees well with observations. In addition to this, we provide Fig. \ref{hist_Q_obs2} to compare the velocities with respect to the centre of mass of the clusters, from which the pre-merger binaries originate. The bimodal distribution in this plot demonstrates that the majority of mergers take place within the mid-plane of the galaxy, where the FS CMa stars cannot be observed due to dust obscuration. Additionally, due to the small number of FS~CMa stars in our sample, detailed statistical analysis cannot be done because the observed data are subject to significant bias.

We conclude that mergers have a significant impact on the population of B-type stars. Despite this, little to no research has focused on these lower-mass stellar mergers. Some dynamical studies have been done, such as the works by \cite{Hurleyetal2001, Hurleyetal2005}. A noteworthy mention is the study of \cite{Leiner2019} that focuses on the population of blue lurkers, the low-luminosity stars that can be found close to the zero-age main sequence. They have a lot in common with the more familiar group of blue straggler stars. They both must have gone through a period of mass transfer, or through a merger or stellar collision event, which caused them to spin up. The fast rotation is what helps the authors to identify candidates for these blue lurkers.

Another notable point of our study is that we detected merger events that occur over a~long period of time and over large distances. Since most of the merger events happen during the CE phase, it is reasonable to assume that a~lot of material is expelled from the system, making a~contribution to the interstellar medium. This may be a relevant consideration, as we can see in the study of young globular clusters with direct N-body modelling done by \cite{Wang2020}, who show the significance of mergers in helping different channels of multiple stellar population formation to work together. 

Our calculations show a high occurrence of low- to intermediate-mass mergers. More than half of the merger products are found in the mass range of the B-type stars. Among B-type stars, the FS~CMa stars exhibit properties that are expected in mergers in certain evolutionary stages. The recent discovery of a post-merger, IRAS 17499+2320 \citep{Korcakova2022}, along with other observed properties of the FS~CMa stars, lead us to the conclusion that the FS~CMa stars are the best group with which to uncover new post-mergers. They are the key to filling the gap in our knowledge of the evolution of stellar mergers, chemical peculiarities of some stars, and stellar magnetic fields.
In a~follow-up paper, we are focusing on the investigation of the distant and very fast merger products as well the overall impact of the low- to intermediate-mass mergers on the chemical enrichment of the Galaxy.

\begin{acknowledgements}
We thank the referee for productive and very helpful comments. They have helped us improve our manuscript significantly.
This work has made use of data from the European Space Agency (ESA) mission {\it Gaia}, processed by the {\it Gaia} DPAC. Part of this research has made use of spectra obtained at the Canada-France-Hawaii Telescope (CFHT), operated by the National Research Council of Canada, the Institut National des Sciences de l’Univers of the Centre National de la Recherche Scientique of France, and the University of Hawaii. These observations were performed with respect for the significant cultural and historic site of Maunakea summit. We thank Ladislav Subr for technical support and computer cluster administration. Part of this research is based on data obtained from the ESO Science Archive Facility with DOIs: \url{https://doi.eso.org/10.18727/archive/24}, \url{https://doi.eso.org/10.18727/archive/50}, \url{https://doi.eso.org/10.18727/archive/71}. Programme IDs of observations made with ESO Telescopes at La Silla and Paranal Observatory are specified in Table~\ref{obsIDpt1}. This work has been supported through the DAAD Bonn-Prague exchange programme. FD acknowledges the support via the Canon Foundation in Europe Fellowship.

\end{acknowledgements}

\bibliographystyle{aa} 
\bibliography{refer} 

\begin{appendix}
\section{W and Q velocities of our simulations}
\begin{figure}[h!]
    \centering
    \includegraphics[width=9cm]{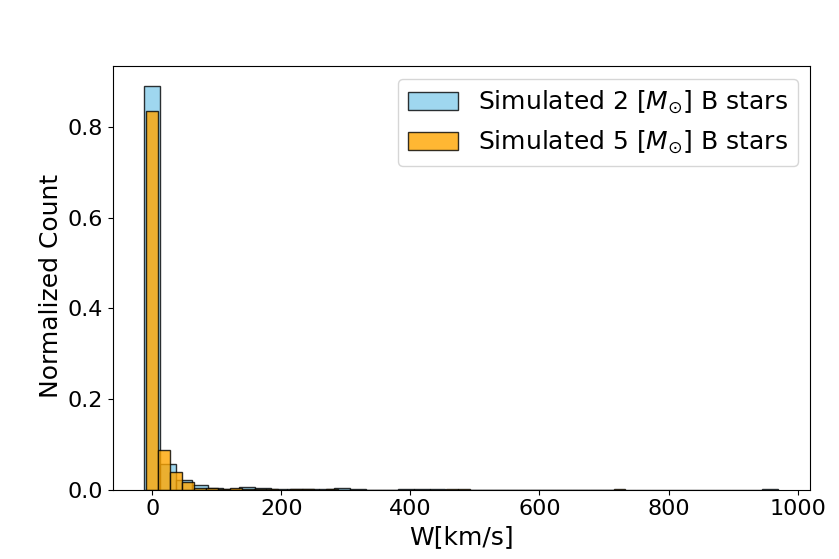}
    \caption{W component of the space velocity relative to the centre of mass of the cluster. Due to the symmetry of the calculated distribution, we show only its positive part of the distribution on the right panel. Both distributions (the 2~$\text{M}_{\odot}$ and 5~$\text{M}_{\odot}$) are normalised in these plots.}
    \label{hist_Q_obs}
\end{figure}


\begin{figure}[h!]
    \centering
    \includegraphics[width=9.5cm]{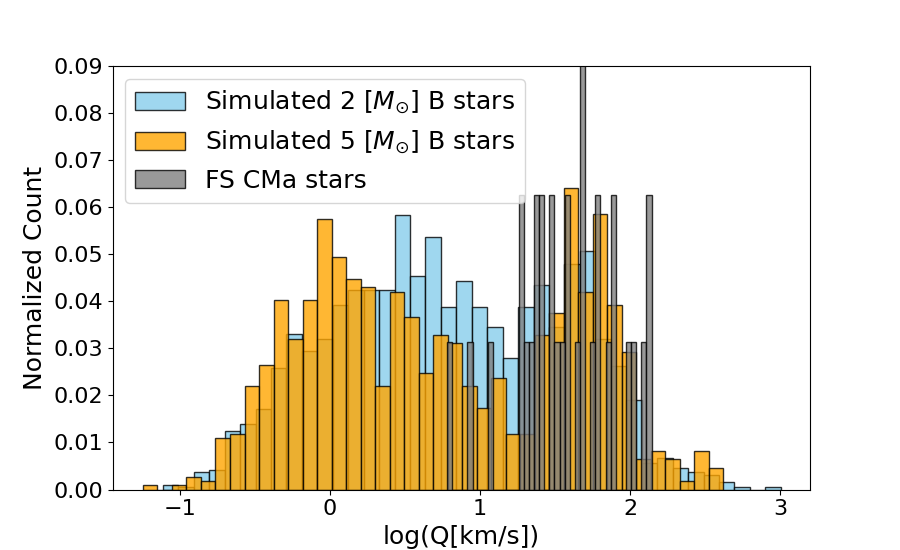}
    \caption{Comparison of the space velocity Q$_{sim}$ of the mergers with respect to the centre of mass of the respective cluster. Only B-type mergers are compared to space velocity Q$_{obs}$ the FS~CMa stars. The histogram is normalised by the total number of points for individual data sets.}
    \label{hist_Q_obs2}
\end{figure}

\section{Tables}

\begin{sidewaystable*}[ht]
\caption{Radial velocities of our sample of FS~CMa stars. }             
\label{velocities}
\centering
\begin{tabular}{l c c c c c c c c}        
\hline\hline                 
IRAS number & Other names & Rv $_{6300}$ [km/s] & Rv $_{6364}$ [km/s] & U [km/s] & V [km/s] & W [km/s] & \textbf{$Q_{obs}$} [km/s]  &   $\text{d}_{GAIA}$ [kpc] \\    
\hline 
03421+2935      &       MWC     728     &       25.62   $\pm$   0.64    &       26.72   $\pm$   0.42    &       -15.47  $\pm$   0.52    &       16.19   $\pm$   0.25    &       -5      $\pm$   0.41    &       22.95   $\pm$   0.72    &       0.3     $\pm$   0.03    \\
03549+5602      &       AS      78      &       -33.9   $\pm$   0.12    &       -33.64  $\pm$   0.03    &       33.4    $\pm$   0.2     &       -10.04  $\pm$   0.16    &       2.78    $\pm$   0.09    &       34.99   $\pm$   0.27    &       2.73    $\pm$   0.11    \\
05111+3244      &       MWC     485     &       6.11    $\pm$   0.02    &       6.34    $\pm$   0.07    &       1.15    $\pm$   0.12    &       -7.69   $\pm$   1.44    &       -17.97  $\pm$   1.59    &       19.58   $\pm$   2.15    &       2.12    $\pm$   0.14    \\
05598-1000      &       AS      116     &       28.24   $\pm$   2.5     &       28.78   $\pm$   1.7     &       -18.65  $\pm$   1.71    &       5.49    $\pm$   1.06    &       -4.57   $\pm$   0.61    &       19.97   $\pm$   2.1     &       0.82    $\pm$   0.01    \\
06070-0938      &       AS      119     &       43.37   $\pm$   0.98    &       44.14   $\pm$   0.99    &       -31.66  $\pm$   1.17    &       -4.3    $\pm$   0.06    &       -2.97   $\pm$   0.17    &       32.08   $\pm$   1.18    &       1.33    $\pm$   0.09    \\
06071+2925      &       PDS     211     &       12.4    $\pm$   1.75    &       --      &       -4.66   $\pm$   0.88    &       3.67    $\pm$   0.2     &       1.98    $\pm$   0.08    &       6.27    $\pm$   0.98    &       1.06    $\pm$   0.02    \\
06158+1517      &       MWC     137     &       41.19   $\pm$   0.93    &       42.11   $\pm$   0.7     &       -29.54  $\pm$   0.51    &       -5.15   $\pm$   1.18    &       -5.81   $\pm$   1.61    &       30.54   $\pm$   2.07    &       5.15    $\pm$   0.68    \\
06259-1301      &       FS      Cma     &       20.91   $\pm$   0.45    &       21.25   $\pm$   0.19    &       -9.15   $\pm$   0.29    &       0.59    $\pm$   0.18    &       7.51    $\pm$   0.07    &       11.85   $\pm$   0.36    &       0.57    $\pm$   0.01    \\
06491-0654      &       HD      50138   &       36.56   $\pm$   3.31    &       37.97   $\pm$   2.13    &       -25.92  $\pm$   2.2     &       -3.07   $\pm$   1.6     &       2.15    $\pm$   0.19    &       26.19   $\pm$   2.73    &       0.35    $\pm$   0.01    \\
06556+1623      &       OY      Gem     &       52.82   $\pm$   0.28    &       53.02   $\pm$   0.4     &       -30.45  $\pm$   3.76    &       -36.47  $\pm$   12.89   &       9.42    $\pm$   1.99    &       48.44   $\pm$   13.58   &       7.44    $\pm$   2.9     \\
07080+0605      &       TYC     175-3772-1      &       9.04    $\pm$   0.82    &       9.39    $\pm$   0.9     &       1.28    $\pm$   0.73    &       3.74    $\pm$   0.52    &       -7.84   $\pm$   0.18    &       8.78    $\pm$   0.91    &       0.56    $\pm$   0.01    \\
07370-2438      &       AS      160     &       55.72   $\pm$   0.33    &       56.14   $\pm$   0.42    &       -62.58  $\pm$   2.39    &       -10.17  $\pm$   0.95    &       -1.49   $\pm$   0.35    &       63.42   $\pm$   2.6     &       3.42    $\pm$   0.17    \\
07377-2523      &       SS      147     &       87.82   $\pm$   0.02    &       88.8    $\pm$   0.19    &       -99.52  $\pm$   3.24    &       -26.93  $\pm$   1.71    &       -16.42  $\pm$   1       &       104.4   $\pm$   3.8     &       3.28    $\pm$   0.16    \\
07418-2850      &       3       Pup     &       24.57   $\pm$   3.75    &       27.52   $\pm$   0.58    &       -46.1   $\pm$   0.93    &       11.66   $\pm$   1.95    &       -14.05  $\pm$   0.1     &       49.6    $\pm$   2.16    &       1.69    $\pm$   0       \\
08128-5000      &       Hen     3-140   &       26.47   $\pm$   0.29    &       26.09   $\pm$   0.30    &       -58.73  $\pm$   0.02    &       -8.34   $\pm$   0.29    &       7.76    $\pm$   0.04    &       59.83   $\pm$   0.29    &       1.4     $\pm$   0.03    \\
09489-6044      &       HD      85567   &       -19.45  $\pm$   1.19    &       -16.45  $\pm$   1.94    &       -45.77  $\pm$   0.34    &       19.85   $\pm$   1.52    &       4.46    $\pm$   0.15    &       50.09   $\pm$   1.57    &       1.05    $\pm$   0.02    \\
12584-4837      &       Hen     3-847   &       -31.45  $\pm$   0.67    &       -31.21  $\pm$   0.02    &       -34.09  $\pm$   2.2     &       27.04   $\pm$   1.35    &       19.18   $\pm$   2       &       47.56   $\pm$   3.28    &       0.63    $\pm$   0.06    \\
17108-3855      &       AS      222     &       -42.29  $\pm$   0.12    &       -42.54  $\pm$   0.16    &       -35.2   $\pm$   0.06    &       11.9    $\pm$   0.95    &       -5.97   $\pm$   1.12    &       37.63   $\pm$   1.47    &       2.53    $\pm$   0.23    \\
17175-3757      &       AS      225     &       -29.32  $\pm$   1.97    &       -28.73  $\pm$   2.01    &       -25.33  $\pm$   1.66    &       -9.52   $\pm$   1.95    &       -3.61   $\pm$   0.61    &       27.3    $\pm$   2.63    &       2.65    $\pm$   0.15    \\
17449+2320      &       StHa    145     &       -14.5   $\pm$   1.37    &       -15.55  $\pm$   1.22    &       -17.02  $\pm$   0.58    &       9.45    $\pm$   0.96    &       14.5    $\pm$   0.71    &       24.28   $\pm$   1.33    &       0.74    $\pm$   0.01    \\
18313-1738      &       MWC     939     &       9.91    $\pm$   0.19    &       10.12   $\pm$   0.73    &       20.28   $\pm$   0.53    &       6.22    $\pm$   0.29    &       -1.19   $\pm$   0.32    &       21.25   $\pm$   0.71    &       0.8     $\pm$   0.03    \\
18316-0028      &       SS      170     &       16.54   $\pm$   0.49    &       16.37   $\pm$   0.73    &       37.8    $\pm$   1.34    &       -3.94   $\pm$   1.08    &       7.67    $\pm$   0.05    &       38.77   $\pm$   1.73    &       2.02    $\pm$   0.11    \\
18406-0508      &       AS      319     &       35.55   $\pm$   0.9     &       36.16   $\pm$   0.77    &       72.12   $\pm$   3.47    &       -31.54  $\pm$   4.88    &       -8.23   $\pm$   1.24    &       79.14   $\pm$   6.11    &       4.79    $\pm$   0.41    \\
19545+3058      &       MWC     623     &       -9.64   $\pm$   0.88    &       -9.21   $\pm$   0.82    &       102.99  $\pm$   5.82    &       -35.58  $\pm$   1.47    &       1.73    $\pm$   0.24    &       108.98  $\pm$   6.01    &       3.7     $\pm$   0.21    \\
20212+3920      &       MWC     342     &       -31.83  $\pm$   1       &       -31.17  $\pm$   0.51    &       55.06   $\pm$   1.28    &       -28.9   $\pm$   0.49    &       -2.14   $\pm$   0.15    &       62.22   $\pm$   1.4     &       1.77    $\pm$   0.04    \\
20493+4849      &       AS      446     &       -26.84  $\pm$   2.86    &       -20.07  $\pm$   1.41    &       71.22   $\pm$   2.44    &       -11.6   $\pm$   2.07    &       -2.55   $\pm$   0.18    &       72.28   $\pm$   3.25    &       3.01    $\pm$   0.11    \\
21095+4726      &       GGR     8       &       -26.66  $\pm$   0.96    &       -27.26  $\pm$   1.02    &       37.82   $\pm$   0.79    &       -13.91  $\pm$   0.98    &       0.99    $\pm$   0.16    &       40.31   $\pm$   1.27    &       2.49    $\pm$   0.07    \\
21263+4927      &       GGR     25      &       -25.17  $\pm$   1.97    &       -23.43  $\pm$   8.71    &       75.37   $\pm$   1.21    &       -8.01   $\pm$   5.4     &       -15.51  $\pm$   0.59    &       77.37   $\pm$   5.68    &       2.45    $\pm$   0.06    \\
21516+5245      &       MWC     645     &       -66.38  $\pm$   0.29    &       -61.57  $\pm$   0.28    &       129.06  $\pm$   17.78   &       -34.7   $\pm$   2.72    &       -13.86  $\pm$   3.43    &       134.38  $\pm$   18.31   &       6.83    $\pm$   1.09    \\
22065+5358      &       MWC     1055    &       -90.66  $\pm$   0.97    &       -90.62  $\pm$   0.34    &       133.64  $\pm$   8.15    &       -55.88  $\pm$   2.15    &       -2.72   $\pm$   0.88    &       144.88  $\pm$   8.48    &       6.34    $\pm$   0.48    \\
22248+6058      &       V669    Cep     &       -21.24  $\pm$   0.62    &       -21.61  $\pm$   0.15    &       23.65   $\pm$   0.38    &       -4.04   $\pm$   0.54    &       -2.18   $\pm$   0.38    &       24.09   $\pm$   0.79    &       0.86    $\pm$   0.05    \\
None    &       FBS0022-021             &       -31.45  $\pm$   0.81    &       -31.61  $\pm$   0.28    &       126.98  $\pm$   16.58   &       -8.58   $\pm$   1.05    &       13.81   $\pm$   3.55    &       128.01  $\pm$   16.99   &       4.2     $\pm$   0.61    \\
\end{tabular}
\tablefoot{We show the results measured by fitting the [O~I] lines $\lambda\lambda$ 6300.304, 6363.776~\AA. The space velocity, Q, was calculated using the averaged values of these measurements. $\text{D}_{GAIA}$ is the distance calculated from the \textit{Gaia} EDR3 parallax.}
\end{sidewaystable*}

\begin{table*}[ht]
\caption{Observations.}             
\label{obsIDpt1}
\centering                          
\begin{tabular}{l c l l c l}        
\hline\hline                 
IRAS number & Other names & Filename & Spectrograph & Date of obs. & Programme/Run ID  \\    
\hline 
03421+2935      &       MWC 728 &       mwc728-1220162i.fits    &       ESPaDOnS        &       2010-08-06      &       10BC12         \\
\hline 
03549+5602      &       AS 78   &       as78-1266547i.fits      &       ESPaDOnS        &       2010-12-18      &       10BC12  \\
        &               &       as78-2372960i.fits      &       ESPaDOnS        &       2019-01-24      &       18BD91         \\
        &               &       as78-2559303i.fits      &       ESPaDOnS        &       2020-12-08      &       20BD94         \\
  \hline 
05111+3244      &       MWC 485 &       mwc485-1260546i.fits    &       ESPaDOnS        &       2010-11-21      &       10BC12         \\
        &               &       mwc485-1266408i.fits    &       ESPaDOnS        &       2010-12-17      &       10BC12         \\
  \hline 
05598-1000      &       AS 116  &       as116-ADP.2014-05-15T14-17-14.497.fits  &       Xshooter        &       2010-02-24      &       084.C-0952(A)   \\
        &               &       as116-1514595iniraf.fits        &       ESPaDOnS        &       2012-01-08      &       11BB05  \\
  \hline 
06070-0938      &       AS 119  &       as119-ADP.2021-02-01T10-11-07.475.fits  &       Uves    &       2020-11-19      &       106.211J.002    \\
        &               &       as119-833562i.fits      &       ESPaDOnS        &       2006-01-13      &       05BD05         \\
        &               &       as119-1267782i.fits     &       ESPaDOnS        &       2010-12-26      &       10BC12         \\
  \hline 
06071+2925      &       PDS211  &       iras06071-1267784i.fits &       ESPaDOnS        &       2010-12-26      &       10BC12         \\
        &               &       iras06071-1524983i.fits &       ESPaDOnS        &       2012-02-13      &       12AC12         \\
        &               &       iras06071-837250i.fits  &       ESPaDOnS        &       2006-02-07      &       06AD02         \\
  \hline 
06158+1517      &       MWC 137 &       mwc137-ADP.2021-02-01T09-55-45.863.fits &       Uves    &       2020-11-09      &       106.21NW.002    \\
        &               &       mwc137-1755718i.fits    &       ESPaDOnS        &       2014-11-10      &       14BC25         \\
        &               &       mwc137-1772737i.fits    &       ESPaDOnS        &       2015-01-08      &       14BC25         \\
  \hline 
06259-1301      &       FS CMa  &       fscma-ADP.2014-05-17T16 39 54.877.fits  &       Xshooter        &       2013-02-14      &       090.D-0212(A)   \\
        &               &       fscma-1835570in.fits    &       ESPaDOnS        &       2015-09-30      &       15BC07         \\
        &               &       fscma-1919610in.fits    &       ESPaDOnS        &       2016-04-16      &       16AC17         \\
        &               &       fscma-2217715in.fits    &       ESPaDOnS        &       2017-11-09      &       17BD96         \\
        &               &       fscma-2335543in.fits    &       ESPaDOnS        &       2018-11-20      &       18BD91         \\
        &               &       fscma-2556563in.fits    &       ESPaDOnS        &       2020-11-28      &       20BD94         \\
        &               &       fscma-833559in.fits     &       ESPaDOnS        &       2006-01-13      &       05BD05         \\
  \hline 
06491-0654      &       HD 50138        &       hd50138-2217763i.fits   &       ESPaDOnS        &       2017-11-09      &       17BD96         \\
        &               &       hd50138-2556567i.fits   &       ESPaDOnS        &       2020-11-28      &       20BD94         \\
  \hline 
06556+1623      &       OY Gem  &       oygem-ADP.2016-09-27T11-56-46.653.fits  &       Feros   &       2014-12-05      &       094.A-9029(D)   \\
        &               &       oygem-ADP.2017-01-05T09-06-13.026.fits  &       Feros   &       2016-12-06      &       098.A-9039(C)   \\
  \hline 
07080+0605      &       TYC 175-3772-1  &       07080-ADP.2016-09-28T06 54 50.729.fits     &       Feros   &       2015-12-07      &       096.A-9030(A)   \\
        &               &       07080-1048977in.fits    &       ESPaDOnS        &       2008-12-20      &       08BD92         \\
        &               &       07080-1266416i.fits     &       ESPaDOnS        &       2010-12-17      &       10BC12         \\
        &               &       07080-1524365i.fits     &       ESPaDOnS        &       2012-02-10      &       12AC12         \\
        &               &       07080-2335545i.fits     &       ESPaDOnS        &       2018-11-20      &       18BD91         \\
        &               &       07080-2559362i.fits     &       ESPaDOnS        &       2020-12-08      &       20BD94         \\
  \hline 
07370-2438      &       AS 160  &       as160cfht1119.fits      &       ESPaDOnS        &       2018-11-19      &               \\
        &               &       as160cfht1221.fits      &       ESPaDOnS        &       2019-12-21      &               \\
        &               &       as160-ADP.2021-02-01T10-25-55.919.fits  &       Uves    &    2020-11-25      &       106.21NW.001    \\
        &               &       as160-ADP.2020-06-26T06-46-20.117.fits  &       Uves    &       2009-02-13      &       082.C-0831(A)   \\
        &               &       as160-1260547i.fits     &       ESPaDOnS        &       2010-11-21      &       10BC12         \\
        &               &       as160-1261667i.fits     &       ESPaDOnS        &       2010-11-25      &       10BC12         \\
        &               &       as160-1524986i.fits     &       ESPaDOnS        &       2016-03-28      &       12AC12         \\
        &               &       as160-1916036i.fits     &       ESPaDOnS        &       2016-03-28      &       16AC17         \\
        &               &       as160-1920104i.fits     &       ESPaDOnS        &       2016-04-18      &       16AC17         \\
        &               &       as160-2335409i.fits     &       ESPaDOnS        &       2018-11-19      &       18BD91         \\
        &               &       as160-2556801i.fits     &       ESPaDOnS        &       2020-11-29      &       20BD94         \\
  \hline 
07377-2523      &       SS147   &       iras07377-1261669i.fits &       ESPaDOnS        &       2010-11-25      &       10BC12         \\
        &               &       iras07377-1266412i.fits &       ESPaDOnS        &       2010-12-17      &       10BC12         \\
  \hline 
07418-2850      &       3Pup    &       3pup-ADP.2016-09-27T09-50-39.525.fits   &       Feros   &       2013-05-09      &       091.D-0221(A)   \\
        &               &       3pup-1059152i.fits      &       ESPaDOnS        &       2009-02-14      &       09AH99         \\
        &               &       3pup-1147172i.fits      &       ESPaDOnS        &       2009-12-09      &       09BC16         \\
        &               &       3pup-1845487i.fits      &       ESPaDOnS        &       2015-11-03      &       15BC04         \\
  \hline 
08128-5000      &       Hen3-140        &       Hen140-ADP.2020-06-26T06-49-12.675.fits &       Uves    &       2009-02-14      &       082.C-0831(A)   \\
\hline 
09489-6044      &       HD 85567        &       hd85567-ADP.2017-06-02T15-33-43.734.fits        &       Xshooter        &       2010-03-06      &       084.C-0952(A)   \\
\hline 
12584-4837      &       Hen 3-847       &       Hen3-847-1523480i.fits  &       ESPaDOnS        &       2012-02-03      &       12AC12         \\
        &               &       Hen3-847-1627161i.fits  &       ESPaDOnS        &       2013-05-18      &       13AC16         \\
  \hline 
17108-3855      &       AS 222  &       as222-1011567i.fits     &       ESPaDOnS        &       2008-07-19      &       08AC06         \\
        &               &       as222-1012421i.fits     &       ESPaDOnS        &       2008-07-25      &       08AC06         \\
        &               &       as222-1546929i.fits     &       ESPaDOnS        &       2012-07-09      &       12AC12         \\
  \hline 
17175-3757      &       AS 225  &       as225-ADP.2020-06-23T08-21-04.621.fits  &       Uves    &       2009-04-19      &       083.C-0676(A)   \\
        &               &       as225-1011921i.fits     &       ESPaDOnS        &       2008-07-21      &       08AC97         \\
        &               &       as225-1546723i.fits     &       ESPaDOnS        &       2012-07-08      &       12AC12         \\
\hline                                   
\end{tabular}
\end{table*}

\begin{table*}[ht]
\caption*{Table \ref{obsIDpt1} continued}             
\label{obsIDpt2}
\centering                          
\begin{tabular}{l c l l c l}        
\hline\hline                 
IRAS number & Other names & Filename & Spectrograph & Date of obs. & Programme/Run ID  \\    
\hline 
17449+2320      &       StHa 145        &       StHa145-ADP.2016-09-28T11-26-16.868.fits        &       Feros   &       2016-04-13      &       097.A-9024(A)   \\
        &               &       StHa145-1012457i.fits   &       ESPaDOnS        &       2008-07-25      &       08AC97         \\
        &               &       StHa145-1525011i.fits   &       ESPaDOnS        &       2012-02-13      &       12AC12         \\
        &               &       StHa145-2005732i.fits   &       ESPaDOnS        &       2016-09-22      &       16BD94         \\
        &               &       StHa145-2169953i.fits   &       ESPaDOnS        &       2017-09-08      &       17BD96         \\
        &               &       StHa145-851537i.fits    &       ESPaDOnS        &       2006-06-09      &       06ad04         \\
        &               &       StHa145-923660i.fits    &       ESPaDOnS        &       2007-07-03      &       07ad02         \\
  \hline 
18313-1738      &       MWC 939 &       mwc939-ADP.2016-09-21T07-46-52.326.fits &       Feros   &       2005-04-20      &       075.D-0177(A)   \\
        &               &       mwc939-20060807-861382i.fits    &       ESPaDOnS        &       2006-08-07      &       06bd02         \\
        &               &       mwc939-ADP.2013-09-27T20-25-49.000.fits &       Uves    &       2009-08-27      &       083.C-0676(A)   \\
        &               &       mwc939-20120707-1546538i.fits   &       ESPaDOnS        &       2012-07-07      &       12AC12         \\
  \hline 
18316-0028      &       SS 170  &       ss170ADP.2020-06-22T05-59-03.934.fits   &       Uves    &       2009-08-08      &       083.C-0676(A)   \\
        &               &       ss170-1220146i.fits     &       ESPaDOnS        &       2010-08-06      &       10BC12         \\
  \hline 
18406-0508      &       AS 319  &       as319-ADP.2020-06-22T06 25 23.708.fits  &       Uves    &       2009-08-27      &       083.C-0676(A)   \\
        &               &       as319-921583i.fits      &       ESPaDOnS        &       2007-06-23      &       07AD02         \\
  \hline 
19545+3058      &       MWC 623 &       mwc623-1850199i.fits    &       ESPaDOnS        &       2015-12-02      &       15BC07         \\
        &               &       mwc623-923680i.fits     &       ESPaDOnS        &       2007-07-03      &       07ad02         \\
  \hline 
20212+3920      &       MWC 342 &       mwc342-1012099i.fits    &       ESPaDOnS        &       2008-07-22      &       08AC97         \\
        &               &       mwc342-1834906i.fits    &       ESPaDOnS        &       2015-09-25      &       15BC07         \\
        &               &       mwc342-1850020i.fits    &       ESPaDOnS        &       2015-12-01      &       15BC07         \\
        &               &       mwc342-808480i.fits     &       ESPaDOnS        &       2005-08-20      &       05BD01         \\
        &               &       mwc342-861548i.fits     &       ESPaDOnS        &       2006-08-09      &       06bd02         \\
        &               &       mwc342-924017i.fits     &       ESPaDOnS        &       2007-07-04      &       07ad02         \\
  \hline 
20493+4849      &       AS 446  &       as446-2007955i.fits     &       ESPaDOnS        &       2016-10-13      &       16BD94         \\
        &               &       as446-2169955i.fits     &       ESPaDOnS        &       2017-09-08      &       17BD96         \\
  \hline 
21095+4726      &       GGR8    &       ggr8-1261815i.fits      &       ESPaDOnS        &       2010-11-26      &       10BC12         \\
        &               &       ggr8-1546727i.fits      &       ESPaDOnS        &       2012-07-08      &       12AC12         \\
        &               &       ggr8-1655852i.fits      &       ESPaDOnS        &       2013-09-23      &       13BC24         \\
  \hline 
21263+4927      &       GGR25   &       ggr25-1260276i.fits     &       ESPaDOnS        &       2010-11-21      &       10BC12         \\
        &               &       ggr25-1546933i.fits     &       ESPaDOnS        &       2012-07-09      &       12AC12         \\
        &               &       ggr25-1654460i.fits     &       ESPaDOnS        &       2013-09-17      &       13BC24         \\
  \hline 
21516+5245      &       MWC 645 &       mwc645-1047424i.fits    &       ESPaDOnS        &       2008-12-10      &       08BD92         \\
        &               &       mwc645-1047563i.fits    &       ESPaDOnS        &       2008-12-11      &       08BD92         \\
  \hline 
22065+5358      &       AS 1055 &       mwc1055-1653969i.fits   &       ESPaDOnS        &       2013-09-15      &       13BC24         \\
        &               &       mwc1055-2008881i.fits   &       ESPaDOnS        &       2016-10-17      &       16BD94         \\
        &               &       mwc1055-2215866i.fits   &       ESPaDOnS        &       2017-10-30      &       17BD96         \\
  \hline 
22248+6058      &       V669 Cep        &       V669Cep-2163529i.fits   &       ESPaDOnS        &       2017-08-14      &       17BD96         \\
        &               &       V669Cep-2163712i.fits   &       ESPaDOnS        &       2017-08-15      &       17BD96         \\
        &               &       V669Cep-2238443i.fits   &       ESPaDOnS        &       2018-01-02      &       17BD96         \\
  \hline 
None    &       FBS0022 021     &       FBS0022-021-833553i.fits        &       ESPaDOnS        &       2006-01-13      &       05BD05         \\
\hline                                   
\end{tabular}
\end{table*}

\end{appendix}

%
%

\end{document}